\documentclass[12pt]{article}

\usepackage[dvips]{epsfig}
\usepackage{amsmath, amsthm, amssymb}
\usepackage{xcolor}
\usepackage{multirow}
\usepackage{graphicx}
\usepackage{subfigure}
\usepackage{epstopdf}
\usepackage{appendix}
\usepackage{listings}
\usepackage{array}
\usepackage{booktabs}
\usepackage{float}

\usepackage{overpic}
\usepackage{supertabular}

\usepackage{float}    % fixed the location to add a figure
\usepackage{setspace}
\usepackage{enumerate}  % list the statements one by one
\usepackage{subfigure}   % combine figures

\usepackage{extarrows}    % add stuff on top on an equal sign

\usepackage[latin1]{inputenc}
\usepackage{tikz}
\usetikzlibrary{shapes,arrows,chains}
\usetikzlibrary{decorations.pathmorphing, backgrounds, positioning, fit, petri, automata}
\usepackage{xcolor}     % define colors
\usepackage{geometry}

\definecolor{yellow1}{rgb}{1,0.8,0.2}      % setting color (without using 'xcolor' package)
\definecolor{LightBlue1}{RGB}{202,225,255}         % setting color, by using 'xcolor' package
\definecolor{SteelBlue3}{RGB}{79,148,205}

\usepackage{moreverb,url}
\usepackage[colorlinks,linkcolor=black,bookmarksopen,bookmarksnumbered,citecolor=black,urlcolor=black]{hyperref}

\newcommand\BibTeX{{\rmfamily B\kern-.05em \textsc{i\kern-.025em b}\kern-.08em
		T\kern-.1667em\lower.7ex\hbox{E}\kern-.125emX}}

\newenvironment{sequation*}{\begin{equation*}\footnotesize}{\end{equation*}}

\newtheorem{theorem}{Theorem}
\newtheorem{lemma}{Lemma}

\begin{document}

\title{\vskip -50pt Detecting the skewness of data from the five-number summary and its application in meta-analysis} %with application to

{\footnotesize
\author{\small Jiandong Shi$^{1}, $ Dehui Luo$^{1}$, Xiang Wan$^{2}, $ Yue Liu$^{3}, $ Jiming Liu$^{4}, $\\
\small Zhaoxiang Bian$^{5}$ and Tiejun Tong$^{1,}$\thanks{Corresponding author. E-mail: tongt@hkbu.edu.hk}\\ \\
{\footnotesize  $^1$Department of Mathematics, Hong Kong Baptist University, Hong Kong}\\
{\footnotesize  $^2$Shenzhen Research Institute of Big Data, Shenzhen, China}\\
{\footnotesize $^3$Cardiovascular Disease Centre, Xiyuan Hospital of China Academy of Chinese Medical Sciences,}\\ {\footnotesize Beijing, China}\\
{\footnotesize $^4$Department of Computer Science, Hong Kong Baptist University, Hong Kong}\\
{\footnotesize $^5$School of Chinese Medicine, Hong Kong Baptist University, Hong Kong}\\
}
}

\date{}
\maketitle

\begin{abstract}
	For clinical studies with continuous outcomes, when the data are potentially skewed,
	researchers may choose to report the whole or part of the five-number summary (the sample median, the first and third quartiles, and the minimum and maximum values)
	rather than the sample mean and standard deviation.
	In the recent literature, it is often suggested to transform the five-number summary back to the sample mean and standard deviation, which can be subsequently used in a meta-analysis.
	However, if a study contains skewed data, this transformation and hence the conclusions from the meta-analysis are unreliable. Therefore, we introduce
	a novel method for detecting the skewness of data using only the five-number summary and the sample size, 
	and meanwhile propose a new flow chart to handle the skewed studies in a different manner. 	
	We further show by simulations that our skewness tests are able to control the type I error rates and provide good statistical power, followed by a simulated meta-analysis and a real data example that illustrate the usefulness of our new method in meta-analysis and evidence-based medicine.
	
	\vskip 12pt
	\noindent
	$Key \ words$: Evidence-based medicine, Five-number summary, Flow chart, Meta-analysis, Skewness test
	
\end{abstract}

\maketitle
\linespread{1.65}
\setlength{\parskip}{0.1\baselineskip}
\baselineskip 20pt
\section{Introduction}\label{sec1}
Meta-analysis is an important tool to synthesize the research findings from multiple studies for decision making.
To conduct a meta-analysis, the summary statistics are routinely collected from each individual study,
and in particular for continuous outcomes, they consist of the sample mean and standard deviation (SD).
In many other studies, if the data are skewed,
researchers may instead report the whole or part of the five-number summary
$\{a,q_1,m,q_3,b\}$, where $a$ is the minimum value, $q_1$ is the first quartile,
$m$ is the sample median, $q_3$ is the third quartile, and $b$ is the maximum value.
More specifically, by letting $n$ be the size of the data, the three common scenarios
for reporting the five-number summary include
\begin{align*}
	\mathcal{S}_1&=\{a,m,b;n\},\\
	\mathcal{S}_2&=\{q_1,m,q_3;n\},\\
	\mathcal{S}_3&=\{a,q_1,m,q_3,b;n\}.
\end{align*}
In practice, however, few existing methods in meta-analysis are able to
pool together the studies with the sample mean and SD
and the studies with the five-number summary.

To overcome this problem, there are two common approaches in the literature.
The first approach is to exclude the studies with the five-number summary from meta-analysis
by labeling them as ``studies with insufficient data".
This approach was, in fact, quite popular in the early years.
Nevertheless, by doing so, valuable information may be excluded
so that the final meta-analytical result can be less reliable or even misleading,
especially when a large proportion of studies are reported with the five-number summary.
In contrast, the second approach is to apply the recently developed methods \cite{hozo2005,Tong2014,luo2016,shi2018}
that convert the five-number summary back to the sample mean and SD,
and then include them in the subsequent meta-analysis.
It is noteworthy that these transformation methods have been attracting increasing attention in meta-analysis and evidence-based practice.
More recently, our transformation methods in Wan et al.\cite{Tong2014}, Luo et al.\cite{luo2016} and Shi et al.\cite{shi2018},
have also been adopted as the default methods for handling the
five-number summary in R packages $meta$ \cite{Balduzzi2019,Schwarzer2022} and $metafor$ \cite{Viechtbauer2010,Viechtbauer2022},
and the three papers have received 4853, 1212 and 154 citations, respectively, in Google Scholar as of 03 March 2023.

Despite the popularity of the second approach, it is also noteworthy that the aforementioned transformation methods are all built on the basis of the normality assumption for the underlying data.
When the data are skewed, however, these normal-based methods may no longer be able to provide reliable estimates for the true sample mean and SD.
For more details, see the motivating examples in Section \ref{section_motivation}.
As a consequence, if we do not handle such skewed studies in a proper way,
it may result in misleading or even completely wrong conclusions in the subsequent meta-analysis \cite{Bono2017,Sun2020}.

This motivates us to perform the normality test for the data first, 
whose result will guide the subsequent steps as presented in the flow chart of Figure \ref{fig:meta_procedure}.

For the normality test, there is a large body of literature on
mainly two different types of tests, (1) the graphical methods \cite{Thode2002}, 
and (2) the quantitative normality test \cite{Dagostino1990,cramer1928,ad1954,shapiro1965,jarque1980}. 
Nevertheless, we note that most existing normality tests require the complete data set
so that they are not applicable when the data include only the whole or part of the five-number summary.
For this issue, Altman and Bland\cite{altman1996} also discussed in their short note as follows:
``When authors present data in the form of a histogram or scatter diagram
then readers can see at a glance whether the distributional assumption is met.
If, however, only summary statistics are presented---as is often the case---this is much more difficult.".

To summarize, when only the five-number summary is available,
there is currently no method available for testing whether the underlying data follow a normal distribution.
In this paper, we propose a skewness test based on the five-number summary
together with the sample size.
Further by the symmetry of the normal distribution,
if the skewness test shows that the data are significantly skewed,
then equivalently we can also conclude that the data are not normally distributed.
For these skewed studies, we provide practitioners with three options in Figure \ref{fig:meta_procedure}.
On the contrary, if the skewness test is not rejected,
then we follow the common practice that assumes the reported data to be normal. Following the above procedure, we will have the capacity to rule out the very skewed studies so that the final meta-analysis can be conducted more reliably than the existing methods in the literature.
Finally, due to the limited information available from the five-number summary,
we believe that our proposed flow chart in Figure \ref{fig:meta_procedure} also provides a reasonable solution for conducting meta-analysis that handles both normal and skewed studies, and we also expect that it may have potential to be widely adopted in meta-analysis and evidence-based practice.

\begin{figure}[H]	
	\centering
	\begin{tikzpicture}[->,>=stealth',shorten >=1pt,auto,node distance=2.4cm,
		thick]
		\node[rectangle,rounded corners,minimum height=6ex,fill=none,draw=SteelBlue3,text=black,text width=30em,text centered]         (test) at (-1.5, 4.5)             {For a study reported with the five-number summary from systematic review, conduct the skewness test under scenario $\mathcal{S}_1$, $\mathcal{S}_2$ or $\mathcal{S}_3$};
		\node[diamond,aspect=4,rounded corners,minimum height=0.8ex,fill=none,draw=SteelBlue3,text=black,text width=7.5em,text centered]         (nrejected) at (-5.2, 1.5)             {Not skewed};
		\node[diamond,aspect=4,rounded corners,minimum height=0.8ex,fill=none,draw=SteelBlue3,text=black,text width=7.5em,text centered]         (rejected) at (2.2, 1.5)             {Skewed};
		\node[rectangle,rounded corners,minimum height=6ex,fill=none,draw=SteelBlue3,text=black,text width=13em]         (out1) at (-5.2, -3)
		{Scenarios $\mathcal{S}_1$ and $\mathcal{S}_2$:\\
			Estimate the sample mean and SD by Luo et al.\cite{luo2016} and Wan et al.\cite{Tong2014}, respectively\\
			
			\

			Scenario $\mathcal{S}_3$:\\
			Estimate the sample mean and SD by Luo et al.\cite{luo2016} and Shi et al.\cite{shi2018}, respectively
		};
		
		\node[rectangle,rounded corners,minimum height=6ex,fill=none,draw=SteelBlue3,text=black,text width=13em]         (out4) at (2.2, -3)
		{1) Exclude the skewed study from the meta-analysis for normal data, or 2) apply the non-normal data transformation methods for skewed studies, or 3) perform a subgroup analysis that separates the normal and skewed studies};
		\node[rectangle,rounded corners,minimum height=6ex,fill=none,draw=SteelBlue3,text=black,text width=28em,text centered]         (Decision) at (-1.5, -7.5)      {Treat the estimated mean and SD  as the true sample values\\ and include the study in the subsequent meta-analysis};
		
		\path %(Start) edge    node {} (test)             % edge[bend left=26]  setting edge not to be a straight line
		(test)  edge    node {} (nrejected)
		(test)  edge    node {} (rejected)
		(nrejected)  edge    node {} (out1)
		(rejected)  edge    node {} (out4)
		(out1)   edge    node {} (Decision);
	\end{tikzpicture}
	\setlength{\abovecaptionskip}{20pt}%
	\setlength{\belowcaptionskip}{0pt}%
	\caption{A flow chart for conducting meta-analysis when some studies from systematic review are reported with the whole or part of the five-number summary.}
	\label{fig:meta_procedure}
\end{figure}

\section{Motivating examples}\label{section_motivation}
To start with, we first present a simulation study to evaluate the performance of the existing transformation methods \cite{Tong2014,luo2016,shi2018} when the underlying distribution is skewed away from normality.
Specifically, we consider four normal-related distributions \cite{forbes2011distributions} as follows:
(i) the skew-normal distribution with parameters $\delta=0$, $\omega=1$ and $\alpha=-10$,
%the exponential distribution with $\lambda=1$, Exponential(1),
(ii) the half-normal distribution with parameters $\mu=0$ and $\sigma^2=1$,
(iii) the log-normal distribution with parameters $\mu=0$ and $\sigma^2=1$,
and (iv) the mixture-normal distribution that takes the values from $N(-2,1)$ with probability 0.3
and from $N(2,1)$ with probability 0.7.
To visualize the skewness of the distributions, the probability density functions of the
four distributions are also plotted in Figure \ref{fig:four_skewed_distribution}. 
It is evident that they are all skewed away, more or less, from the normal distribution.

\begin{table}[h]
	\setlength{\abovecaptionskip}{0pt}%
	\setlength{\belowcaptionskip}{6pt}%
	\caption{The true and estimated averages (standard errors) of the sample mean and SD under scenario $\mathcal{S}_1$ for the four normal-related distributions.}\label{motivation_simulation}
	\centering
	\begin{tabular}{c|cc|cc}
		\hline
		\multirow{2}{*}{Distribution}& \multicolumn{2}{c|}{Sample mean}  &\multicolumn{2}{c}{Sample SD}\\
		&True value &Estimated value \cite{luo2016} &True value &Estimated value \cite{Tong2014}\\
		\hline
		$\text{Skew-normal}$        &-0.79 (0.04)  &-0.73 (0.05)  &0.61 (0.04) &0.57 (0.07) \\
		$\text{Half-normal}$            &0.80  (0.04)  &0.73  (0.05)  &0.60 (0.04) &0.54 (0.07) \\
		$\text{Log-normal}$             &1.65  (0.16)  &1.53  (0.32)  &2.09 (0.56) &3.10 (1.57) \\
		$\text{Mixture-normal}$      &0.80  (0.15)  &1.34  (0.14)  &2.09 (0.08) &1.63 (0.11) \\
		\hline
	\end{tabular}
\end{table}

\begin{figure}[h]
	\centering
	\includegraphics[width=0.95\textwidth]{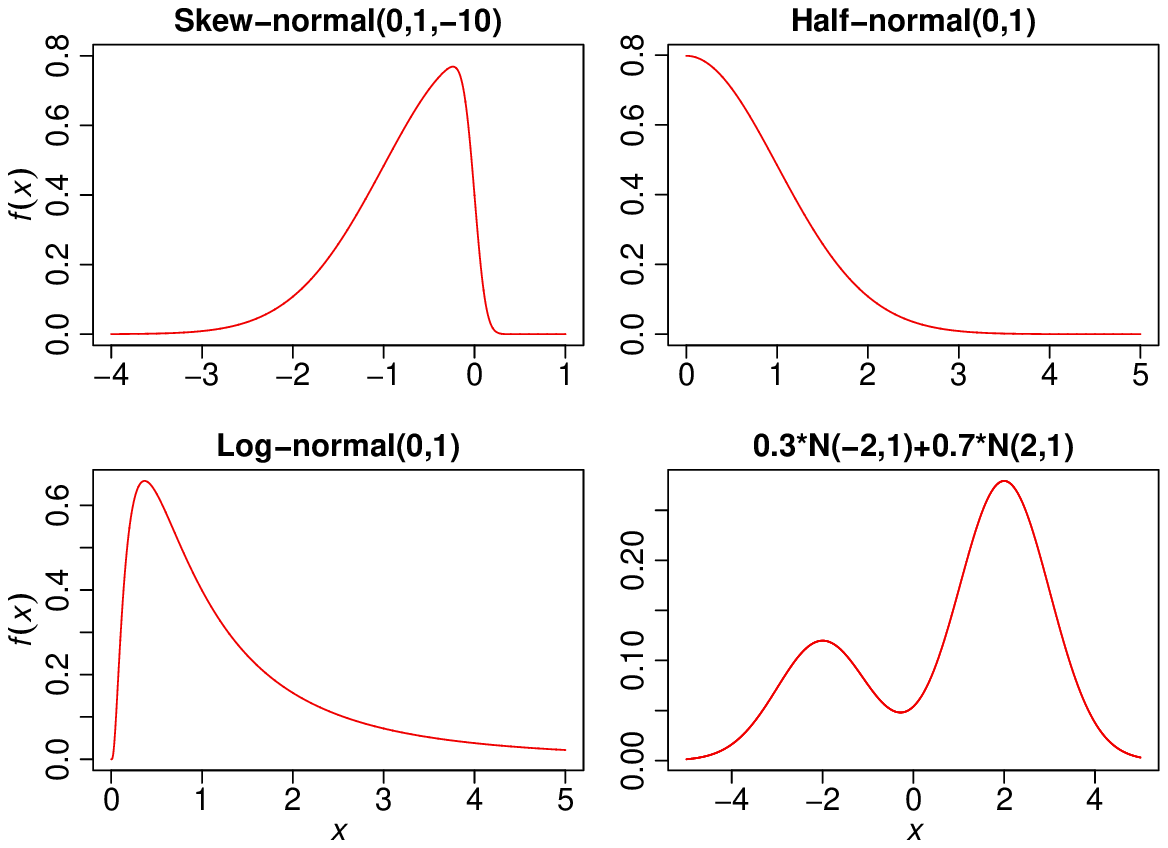}
	\setlength{\abovecaptionskip}{0pt}%
	\setlength{\belowcaptionskip}{0pt}%
	\caption{Probability density functions of the four normal-related distributions
		including Skew-normal(0, 1,-10), Half-normal(0, 1), Log-normal(0, 1), and
		0.3*$N$(-2, 1)+0.7*$N$(2, 1).}\label{fig:four_skewed_distribution}
\end{figure}

Next, for each distribution, a sample of size 200 is randomly generated.
With the complete sample, we can readily compute the sample mean and SD,
and also collect the sample median, the minimum and maximum values.
Now to evaluate the normal-based methods for transformation, we further apply Luo et al.\cite{luo2016} to estimate the sample mean and Wan et al.\cite{Tong2014} to estimate the sample SD under scenario $\mathcal{S}_1$.
With 100,000 simulations,
we report the averages (standard errors) of the estimated sample mean and SD, together with the averages (standard errors) of the true sample mean and SD,
in Table \ref{motivation_simulation}.
From the simulated results, it is evident that
the converted sample mean and SD using the normal-based methods are less accurate for all four skewed distributions.
In particular, we note that the sample SD is significantly overestimated for $\text{Log-normal}(0,1)$, and the sample mean is significantly overestimated for the mixture-normal distribution.

\begin{table}[h]
	\centering    
	\setlength{\abovecaptionskip}{2pt}%
	\setlength{\belowcaptionskip}{10pt}%
	\caption{The summary statistics, sample median (IQR) [sample size], of the four studies in the meta-analysis from Wu and Yang.\cite{wuyang2020}}\label{source_data}
	\begin{tabular}{c|c|c}
		\hline
		~~~~~~~~~~~Study~~~~~~~~~~               &~~~~~~~~~~~Nonsurvivors~~~~~~~~~~~                   &~~~~~~~~~~~Survivors~~~~~~~~~~\\
		\hline
		Chen et al.\cite{chen2020}      &28 (18-47) [113]	    &20 (14.8-32) [161] \\
		
		Du et al.\cite{du2020}	   &27 (20-37) [21]	    &22 (14-40.5) [158] \\
		
		Wang et al.\cite{wang2020}     &24 (19-49) [65]            &28 (17-43) [274] \\
		
		Zhou et al.\cite{zhou2020} &40 (24-51) [54]         &27 (15-40) [135]\\
		\hline
	\end{tabular}
\end{table}

Our second example is a real study that investigates the impact of COVID-19 on liver dysfunction by a meta-analysis \cite{wuyang2020}.
The serum alanine aminotransferase (ALT), as an important index to measure the
dysfunction of the liver, was a primary outcome of interest.
By setting the nonsurvivors and survivors as the case and control groups,
the liver dysfunction can be compared by the ALT level difference between the two groups.
Four clinical studies that paid attention to the ALT level were included in the meta-analysis
with the sample median and the interquartile range (IQR) being reported in Table \ref{source_data}.
The potential skewness of the underlying data
can be observed by comparing the distances between the sample median and the first quartile or the third quartile.
Taking the nonsurvivors group in Wang et al.\cite{wang2020} as an example,
the distance between the sample median to the third quartile $(49-24=25)$
is five times as that between the sample median to the first quartile $(24-19=5)$,
indicating a large degree of skewness.
For more details, see Section \ref{section_realdata}
where the skewed groups with statistical significance are all identified.
Such skewed data, if not properly handled, may
lead to unreliable or even misleading conclusions for decision making in evidence-based practice.

\section{Detecting the skewness from the five-number summary}\label{main}
As sketched in Figure \ref{fig:meta_procedure}, to handle the clinical studies reported with the whole or part of the five-number summary,
the first and foremost thing is to detect whether or not the data follow a normal distribution.
When the normality assumption does not hold, the reported data from the clinical study were often skewed,
which is, in fact, one main reason why researchers had preferred to report the five-number summary.
In this section, we will formulate the null and alternative hypotheses for detecting the skewness of data under the three common scenarios,
and then construct their test statistics,
as well as derive their null distributions and the critical regions.

Let $X_{1}, X_{2}, \ldots, X_{n}$ be a random sample of size $n$
from the normal distribution with mean $\mu$ and variance $\sigma^2$,
and $X_{(1)}\leq X_{(2)}\leq \cdots \leq X_{(n)}$ be the corresponding order statistics.
Then, for simplicity, by letting $Q$ be a positive integer,
the five-number summary can be represented as
$a=X_{(1)}$, $q_1=X_{(Q+1)}$, $m=X_{(2Q+1)}$, $q_3=X_{(3Q+1)}$ and $b=X_{(n)}$.
Let also $X_{i}=\mu+\sigma Z_{i}$ for $i=1, \ldots, n$, or equivalently,
\begin{equation}\label{base}
	X_{(i)}=\mu+\sigma Z_{(i)},
\end{equation}
where $Z_{1}, Z_{2}, \ldots,Z_{n}$ are independent random variables from the standard normal distribution,
and $Z_{(1)}\leq Z_{(2)}\leq \cdots \leq Z_{(n)}$ are the order statistics.
Lastly, when $Q$ is not an integer, we suggest to apply the interpolation method
to calculate the critical values with details in Appendix B.

\subsection{Detecting the skewness under scenario $\mathcal{S}_{1}=\{a,m,b;n\}$}\label{scenario1}
We first consider scenario $\mathcal{S}_1$ where the minimum, median, and maximum values are available together with the sample size.
When the data are normally distributed,
we expect that the distance between $a$ and $m$ should be not far away from the distance between $m$ and $b$.
More specifically, by Lemma \ref{lemma1} in Appendix A and the facts that $E(m-a)=\sigma E(Z_{(2Q+1)}-Z_{(1)})$ and $E(b-m)=\sigma E(Z_{(n)}-Z_{(2Q+1)})$, we have $E(m-a)=E(b-m)$.
In view of this, we define $\theta_1=E(b-m)-E(m-a) = E(a+b-2m)$ as the level of skewness
for the underlying distribution of the data.
Then to detect the skewness of data,
we propose to consider the following hypotheses:
\begin{equation*}\label{hypothesis_s1}
	H_0: \theta_1=0 \quad\quad \text{versus} \quad\quad H_1:\theta_1 \neq 0.
\end{equation*}
If the null hypothesis is rejected, we then conclude that the data are significantly skewed,
and moreover by the flow chart in Figure \ref{fig:meta_procedure},
we recommend practitioners to take the proper choice from the three options
for skewed studies.

Now to test whether $\theta_1=0$ under scenario $\mathcal{S}_1$, by the Wald test \cite{Casella2002},
we consider the test statistic as
\begin{equation*}
	W_1=\dfrac{a+b-2m}{{\rm SE}(a+b-2m)},
\end{equation*}
where ${\rm SE}(a+b-2m)$ denotes the standard error of $a+b-2m$ under the null hypothesis.
By formula (\ref{base}), we can rewrite ${\rm SE}(a+b-2m)=\sigma \delta_1(n)$,
where $\sigma$ is the SD of the normal distribution
and $\delta_1(n)={\rm SE}(Z_{(1)}+Z_{(n)}-2Z_{(2Q+1)})$.
Next, for the unknown $\sigma$, we consider to estimate it by the method in Wan et al.\cite{Tong2014}
Specifically, we have $\hat{\sigma}_1=(b-a)/\xi(n)$, where $\xi(n)=2\Phi^{-1}[(n-0.375)/(n+0.25)]$
and $\Phi^{-1}$ is the quantile function of the standard normal distribution.
Finally, by noting that $\delta_1(n)$ and $\xi(n)$ are fixed values for any given $n$,
we remove them from the Wald statistic and that yields our final test statistic as
\begin{equation}\label{test_statistic_s1}
	T_1=\dfrac{a+b-2m}{b-a}.
\end{equation}
In the special case when $a=b$, all the observations are tied 
so that a test for skewness may not be possible.
To further derive the null distribution of the test statistic $T_1$, 
we consider two different approaches where the first one is to derive the asymptotic null distribution when $n$ tends to infinity,
and the second one is to derive the exact null distribution for any fixed $n$.

For the first approach, noting that $T_1$ involves the extreme order statistics, 
the asymptotic null distribution will not follow a normal distribution
as that for a classical Wald statistic \cite{Ferguson1996}.
To further clarify it, when $n$ is large,
the extreme order statistics will tend to be less stable than the intermediate order statistics
and hence provide a slower convergence rate toward the asymptotic distribution for the given  test statistic.
Specifically by Theorem \ref{asy_s1} in Appendix A, under the null hypothesis, we show that
\begin{equation}\label{t1_convergency}
	\sqrt{2\ln(n)}\xi(n)T_1 \overset{D}{\longrightarrow} {\rm Logistic}(0,1), \quad {\rm as}\ n\rightarrow \infty,
\end{equation}
where $\overset{D}{\longrightarrow}$ denotes the convergence in distribution,
and Logistic$(0,1)$ represents the logistic distribution
with location parameter $\mu=0$ and scale parameter $s=1$.
Noting also that the asymptotic null distribution is symmetric about zero,
we can specify the critical region of size $\alpha$ as
$\{t_{1,{\rm obs}}: |t_{1,{\rm obs}}|> l_{\alpha/2}/(\sqrt{2\ln(n)}\xi(n))\},$
where $t_{1,{\rm obs}}$ is the observed value of $T_1$
and $l_{\alpha/2}$ is the upper $\alpha/2$ quantile of ${\rm Logistic}(0,1)$.
Despite of the elegant analytical results,
the asymptotic test by (\ref{t1_convergency}) will have a serious limitation
that the convergence rate is relatively slow at the order of $\sqrt{2\ln(n)}\xi(n)$.
Moreover, the simulation results in Section \ref{simulation}
will show that the asymptotic null distribution fails to control the type I error rates
for some small sample sizes.

To improve the detection accuracy, our second approach is to derive the exact null distribution of $T_1$ for any fixed $n$.
By (\ref{base}) and (\ref{test_statistic_s1}), we can represent the test statistic as
\begin{equation}\label{sampling_s1}
	T_1=\frac{Z_{(1)}+Z_{(n)}-2Z_{(2Q+1)}}{Z_{(n)}-Z_{(1)}}.
\end{equation}
Since the right-hand side of (\ref{sampling_s1}) is purely a function of the order statistics of the standard normal distribution,
the null distribution of $T_1$ will be free of the parameters $\mu$ and $\sigma^2$.
Moreover, we have derived the sampling distribution of $T_1$ under the null hypothesis in Theorem \ref{sampling_distribution_s1} of Appendix A.
Further by the symmetry of the null distribution,
the critical region of size $\alpha$ can be specified as
\begin{equation*}
	\{t_{1,{\rm obs}}:|t_{1,{\rm obs}}|> c_{1,\alpha/2}(n)\},
\end{equation*}
where $c_{1,\alpha/2}(n)$ is the upper $\alpha/2$ quantile of the null distribution of $T_1$
for the sample size $n$.
If the test is rejected based on the reported summary statistics,
we then conclude that the data from the study are significantly skewed.

From the practical point of view, however, the null distribution of $T_1$ has a complicated
form so that the true values of $c_{1,\alpha/2}(n)$ may not be readily known.
To help practitioners and also promote the new test,
by the R software we have provided the numerical values of $c_{1,\alpha/2}(n)$
for $n$ up to 401 with $\alpha=0.05$ in Table \ref{c1_table} of Appendix B.
Moreover, an approximate formula $c_{1,0.025}(n) \approx 1/\ln (n+9)+2.5/(n+1)$ is also given for easy implementation of the critical values for any given sample size.
It is evident, as shown in Figure \ref{fig:c_combine} of Appendix B,
that the approximation is quite accurate so that it can serve well as a ``rule of thumb" for practical use.
Specifically by the rule of thumb, the skewness test can be performed
by first computing the absolute value of the observed test statistic,
and then examining whether it is larger or smaller than the approximated threshold value at $1/\ln(n+9)+2.5/(n+1)$.

\subsection{Detecting the skewness under scenario $\mathcal{S}_{2}=\{q_1,m,q_3;n\}$}\label{scenario2}
Under scenario $\mathcal{S}_{2}$, the reported summary data include
the first quartile, the median and the third quartile together with the sample size.
When the data are normally distributed, we expect that the distance between $q_1$ and $m$
should be close to the distance between $m$ and $q_3$.
Specifically, by Lemma \ref{lemma1} in Appendix A and the facts that $E(m-q_1)=\sigma E(Z_{(2Q+1)}-Z_{(Q+1)})$ and $E(q_3-m)=\sigma E(Z_{(3Q+1)}-Z_{(2Q+1)})$, we have $E(m-q_1)=E(q_3-m)$.
We then define $\theta_2=E(q_3-m)-E(m-q_1) = E(q_1+q_3-2m)$ as the level of skewness
for the underlying distribution of the data.
Finally for detecting the skewness of data, we consider the following hypotheses:
\begin{equation*}\label{hypothesis_s2}
	H_0: \theta_2=0 \quad\quad \text{versus} \quad\quad H_1:\theta_2 \neq 0.
\end{equation*}
If the null hypothesis is rejected, we conclude
that the underlying distribution of the data is significantly skewed.

Following the same spirit as under scenario $\mathcal{S}_1$,
for the above hypotheses we consider the test statistic
\begin{equation}\label{test_statistic_s2}
	T_2=\frac{q_{1}+q_{3}-2m}{q_3-q_1},
\end{equation}
\noindent
where $q_3> q_1$.
Note that $T_2$ has also been adopted by Groeneveld and Meeden\cite{Groeneveld1984skewness}
as a measure of skewness.
Moreover, unlike the test statistic $T_1$ that involves the extreme order statistics,
the asymptotic normality of $T_2$ can be readily established.
Specifically in Theorem \ref{asy_s2} of Appendix A, we have shown that
\begin{equation}\label{t2_convergency}
	0.74\sqrt{n}T_2 \overset{D}{\longrightarrow} N(0,1), \quad {\rm as}\ n\rightarrow \infty.
\end{equation}
Further by (\ref{t2_convergency}), the critical region of size $\alpha$ can be approximately as
$\{t_{2,{\rm obs}}: |t_{2,{\rm obs}}|> z_{\alpha/2}/(0.74\sqrt{n})\},$
where $t_{2, {\rm obs}}$ is the observed value of $T_2$
and $z_{\alpha/2}$ is the upper $\alpha/2$ quantile of the standard normal distribution.

Nevertheless, given that the asymptotic critical values can be quite large especially for small sample sizes,
the above asymptotic test may not provide an adequate power for detecting the skewness.
To further improve the detection accuracy,
we have also derived the exact null distribution of $T_2$
in Theorem \ref{sampling_distribution_s2} of Appendix A for any fixed $n$.
Noting also that the null distribution of $T_2$ is symmetric about zero,
we can specify the exact critical region of size $\alpha$ as follows:
\begin{equation*}
	\{t_{2,{\rm obs}}:|t_{2,{\rm obs}}|> c_{2,\alpha/2}(n)\},
\end{equation*}
\noindent
where $c_{2,\alpha/2}(n)$ is the upper $\alpha/2$ quantile of the null distribution of $T_2$
for the sample size $n$.
If the observed value of $T_2$ falls in the critical region,
it is concluded that the data are significantly skewed away from normality.

Finally, as that for scenario $\mathcal{S}_1$, we note that obtaining the critical values by Theorem \ref{sampling_distribution_s2} in Appendix A is rather complicated and not readily accessible.
Thus to help practitioners,
we have computed the numerical values of $c_{2,\alpha/2}(n)$ for $\alpha=0.05$ with $n$ up to 401 in Table \ref{c2_table} of Appendix B.
For ease of implementation, an approximate formula for the critical values is also provided as $c_{2,0.025}\approx 2.65/\sqrt{n}-6/n^{2}$,
with its approximation accuracy reported in Figure \ref{fig:c_combine} of Appendix B.
Consequently,
it also provides a convenient way for practitioners to detect the skewness of data by hand.

%---------- Subsection of S3 -----------------
\subsection{Detecting the skewness under scenario $\mathcal{S}_{3}=\{a,q_1,m,q_3,b;n\}$}
In this section, we consider the detection of skewness under scenario $\mathcal{S}_3$
when the five-number summary is fully available together with the sample size.
For normal data, we have $E(m-a)=E(b-m)$ and $E(m-q_1)=E(q_3-m)$, or equivalently,
$\theta_1=E(a+b-2m)=0$ and $\theta_2=E(q_1+q_3-2m)=0$.
Noting also that the summary data under scenario $\mathcal{S}_3$ is the union of
those under scenarios $\mathcal{S}_1$ and $\mathcal{S}_2$,
we consider the following joint hypothesis for detecting the skewness of data:
\begin{equation*}\label{hypothesis_s3}
	H_0: \theta_1=0 \ {\rm and}\ \theta_2=0 \quad \text{versus} \quad H_1:\theta_1\neq 0 \ {\rm or}\ \theta_2\neq 0.
\end{equation*}
If the joint null hypothesis is rejected, then the data will be claimed as significantly skewed,
either in the intermediate region or in the tail region of the underlying distribution.

Following the similar arguments as under scenarios $\mathcal{S}_1$ and $\mathcal{S}_2$,
$\hat{\theta}_1=a+b-2m$ is the sample estimate of the skewness $\theta_1$,
and $\hat{\theta}_2=q_1+q_2-2m$ is the sample estimate of the skewness $\theta_2$.
Thus to test the joint null that $\theta_1=0$ and $\theta_2=0$,
we follow the analysis of variance (ANOVA) 
and take the maximum of their absolute sample estimates as the test statistic \cite{Casella2002}.
Specifically, if the maximum value is larger than a given threshold,
then the test will be rejected so that either $\theta_1$ or $\theta_2$ will be concluded as nonzero.
Meanwhile, to make the two test components comparable,
we also standardize them and yield the test statistic as
$W_3=\max\{|W_1|,|W_2|\}$, where $W_1=(a+b-2m)/{\rm SE}(a+b-2m)$ as already defined
and $W_2=(q_1+q_3-2m)/{\rm SE}(q_1+q_3-2m)$.
Further by Theorem \ref{k_n} in Appendix A,
we replace ${\rm SE}(a+b-2m)$ and ${\rm SE}(q_1+q_3-2m)$ by their respective estimates,
and formulate the final test statistic as
\begin{equation}\label{test_statistic_s3}
	T_3=\max\left\{\dfrac{2.65\ln (0.6n)}{\sqrt{n}}\bigg|\dfrac{a+b-2m}{b-a}\bigg|, \ \bigg|\dfrac{q_1+q_3-2m}{q_3-q_1}\bigg|\right\},
\end{equation}
where $b\geq q_3 > q_1\geq a$.
In addition, by (\ref{test_statistic_s1}), (\ref{test_statistic_s2})
and the fact that the weight to the first component is purely a function of $n$,
it follows that $T_3$ will be independent of $\mu$ and $\sigma^2$ under the joint null hypothesis.
Then accordingly, we can propose the critical region of size $\alpha$ as
\begin{equation*}
	\{t_{3,{\rm obs}}:t_{3,{\rm obs}}> c_{3,\alpha}(n)\},
\end{equation*}
where $t_{3,{\rm obs}}$ is the observed value of $T_3$,
and $c_{3,\alpha}(n)$ is the upper $\alpha$ quantile of its null distribution for the sample size $n$.

The same as before, we also apply the sampling method
to numerically compute the critical values of $c_{3,\alpha}(n)$ for $\alpha=0.05$ with $n$ up to 401, and present them in Table \ref{c3_table} of Appendix B.
Moreover, we provide an approximate formula for the critical values as $c_{3,0.05}(n)\approx 3/\sqrt{n}-40/n^3$ and its approximation accuracy is shown in Figure \ref{fig:c_combine} of Appendix B. Consequently, it readily provides an alternative way to detect the skewness of the reported data.

\begin{figure}[h]
	\centering
	\includegraphics[width=0.9\textwidth]{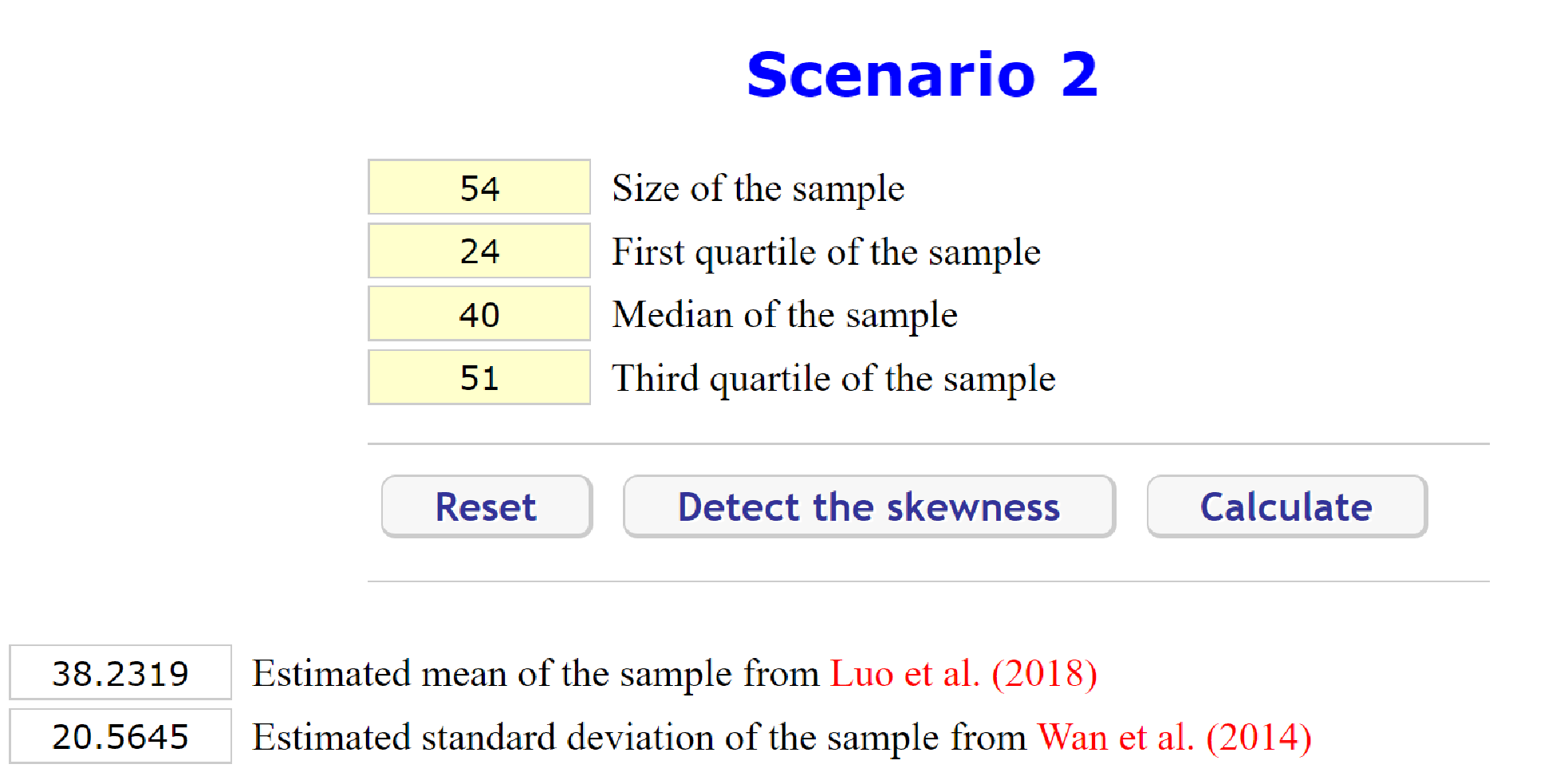}
	\setlength{\abovecaptionskip}{10pt}%
	\setlength{\belowcaptionskip}{0pt}%
	\caption{An example for implementing the online calculator under scenario $S_2$, with the sample data from the nonsurvivors group in Zhou et al.\cite{zhou2020}}
	\label{fig:onlinecalculator}
\end{figure}

Last but not least, we have also launched an online calculator for practitioners to implement the flow chart including the skewness test and the data transformation at
\url{http://www.math.hkbu.edu.hk/~tongt/papers/median2mean.html}.
Our online calculator is very user-friendly, and for illustration,
we consider scenario $\mathcal{S}_2$ with the reported data  $\{q_1,m,q_3;n\}=\{24,40,51;54\}$ from the nonsurvivors group in Zhou et al.\cite{zhou2020}
As shown in Figure \ref{fig:onlinecalculator},
with the summary data in the corresponding entries,
one can click on the {\it Detect the skewness} button to examine whether the data are skewed away from normality.
A popup window will then appear showing the test result,
and specifically for the given data, there is no significant evidence to show
that the data are skewed.
In view of this, we further click on the {\it Calculate} button to perform data transformation by the normal-based methods, which yields the sample mean estimate as 38.2319 by Luo et al.\cite{luo2016} and
the sample SD estimate as 20.5645 by Wan et al.\cite{Tong2014}

\section{Simulation studies}\label{simulation}
In this section, we conduct simulation studies
to evaluate the performance of the three skewness tests.
We first assess the type I error rates of the new tests with the asymptotic,
exact, and approximated critical values at the significance level of 0.05,
and then compute and compare their statistical power under four skewed alternative distributions.
Moreover, we also conduct a simulated meta-analysis
to demonstrate the usefulness of the skewness tests in practice.

\subsection{Type I error rates}\label{t1e}
\begin{figure}[H]
	\centering
	\includegraphics[width=0.75\textwidth]{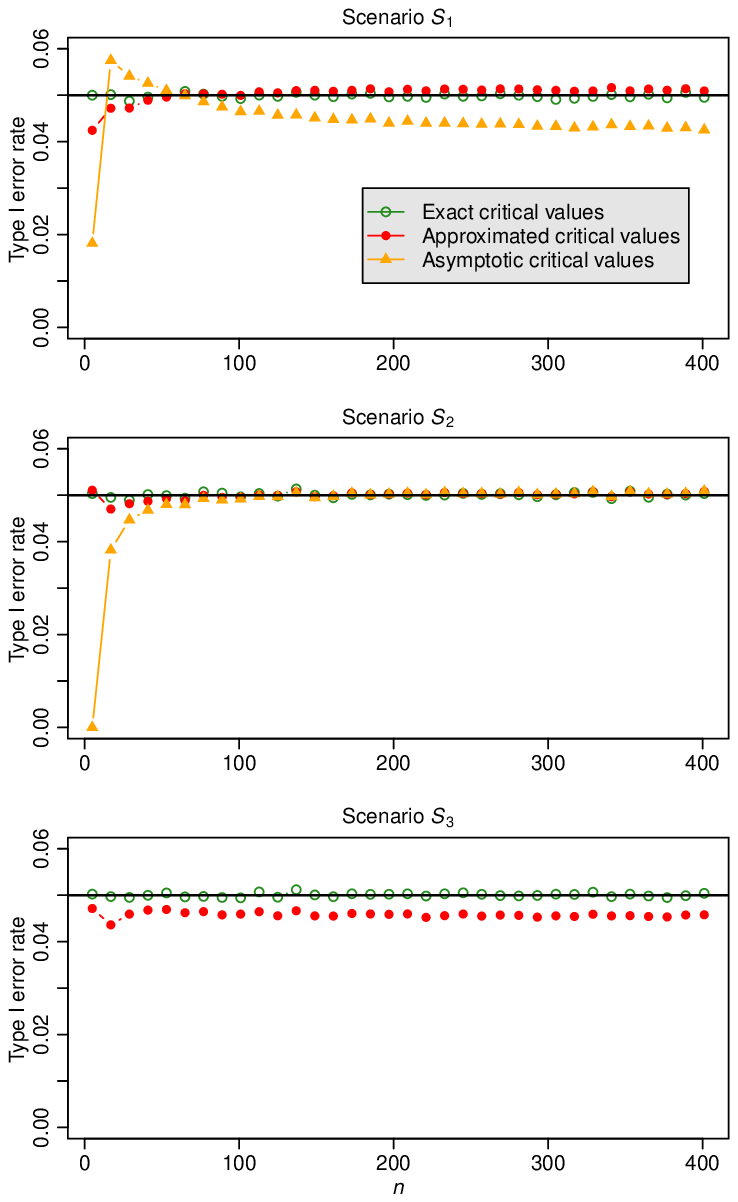}%
	\setlength{\abovecaptionskip}{10pt}%
	\setlength{\belowcaptionskip}{0pt}%
	\caption{The type I error rates for the proposed test statistics under the three scenarios for $n$ up to 401. The solid orange triangles represent the tests with the asymptotic critical values, the empty green points represent the tests with the exact critical values, and the solid red points represent the tests with the approximated critical values.}
	\label{fig:type1error}
\end{figure}

To examine whether the type I error rates are well controlled,
we first generate a sample of size $n$ from the null distribution,
and without loss of generality, we consider the standard normal distribution.
Then for the proposed test under scenario $\mathcal{S}_1$,
we record the summary statistics $\{a , m, b \}$ from the simulated sample,
and compute the observed value of the test statistic $t_{1,{\rm obs}}$ by (\ref{test_statistic_s1}).
Further by comparing $t_{1,{\rm obs}}$ with the asymptotic, exact, and approximated critical values respectively,
we can make a decision whether to reject the null hypothesis, or equivalently,
whether a type I error will be made.
Finally, we repeat the above procedure for 1,000,000 times with $n$ ranging from 5 to 401,
compute the type I error rates for the three different tests
as reported in Figure \ref{fig:type1error}.

It is evident from the simulated results that, under scenario $\mathcal{S}_1$,
the proposed tests with the exact and approximated critical values
perform nearly the same, and they both control the type I error rates at the significance level of 0.05 regardless of the sample size.

\noindent
This, from another perspective, demonstrates that our approximate formula of the critical values
is rather accurate and can be recommended for practical use.
In contrast, for the asymptotic test with the null distribution specified in (\ref{t1_convergency}),
the type I error rates are less well controlled, either inflated or too conservative.
For example, the type I error rate is as high as 0.057 when $n=17$,
and it is always less than 0.05 when $n$ is large,
even though the simulated type I error rate does converge to the significance level as $n$ goes to infinity
which coincides with the theoretical result of Theorem \ref{asy_s1} in Appendix A.

\begin{figure}[h]
	\centering
	\includegraphics[width=0.9\textwidth]{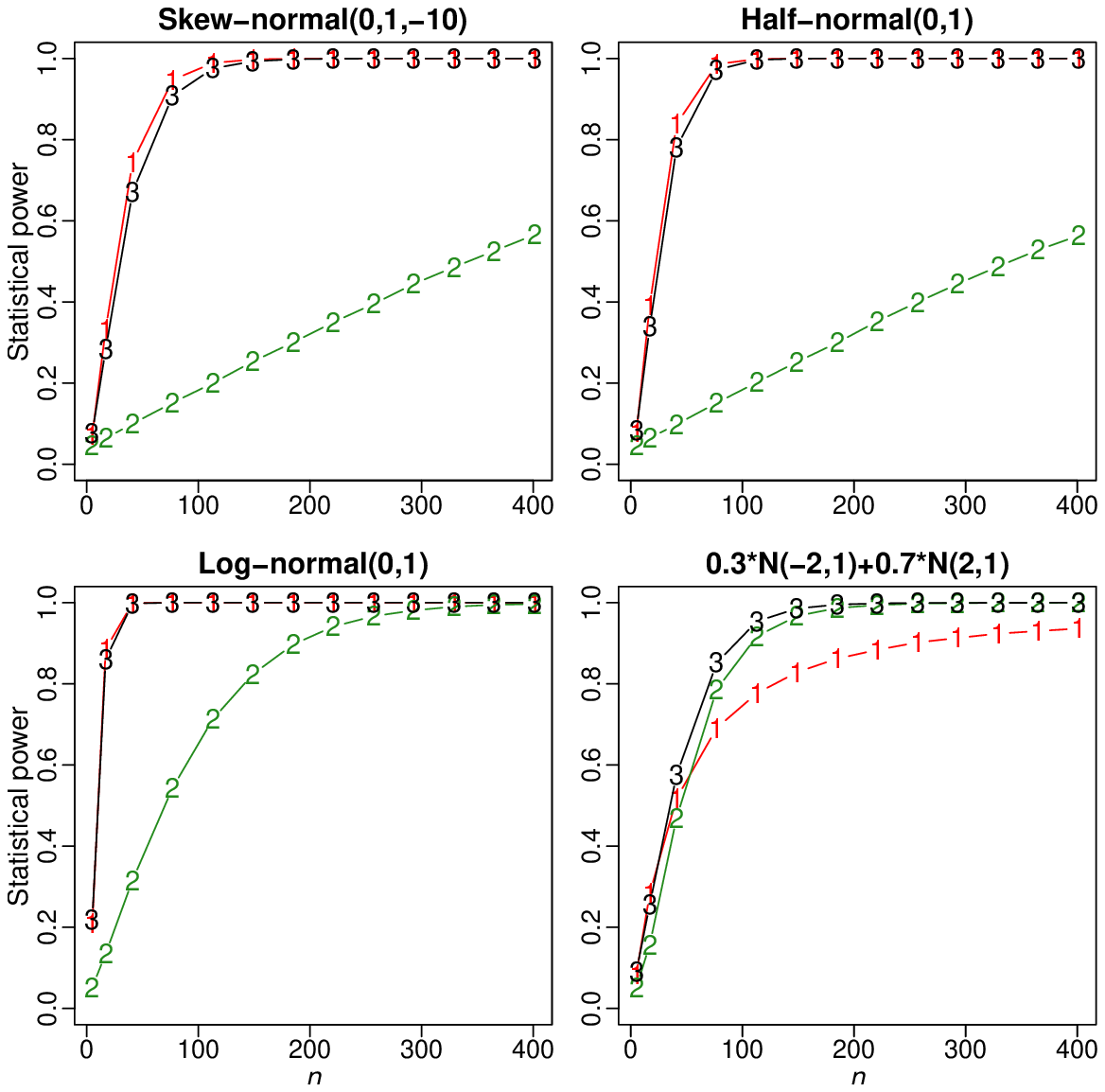}%
	\setlength{\abovecaptionskip}{0pt}%
	\setlength{\belowcaptionskip}{0pt}%
	\caption{The statistical power of the proposed tests under the three scenarios for $n$ up to 401.
		``1" represents the statistical power under scenario $\mathcal{S}_1$,
		``2" represents the statistical power under scenario $\mathcal{S}_2$,
		and ``3" represents the statistical power under scenario $\mathcal{S}_3$.}
	\label{fig:power}
\end{figure}

To assess the type I error rates under the last two scenarios,
we record instead the summary statistics $\{q_1,m,q_3\}$ or $\{a,q_1,m,q_3,b\}$ from the simulated sample,
compute the observed value of the test statistic by (\ref{test_statistic_s2}) or (\ref{test_statistic_s3}),
and then compare it with the different
critical values to determine whether the null hypothesis will be rejected.
Under scenario $\mathcal{S}_2$, the tests with the exact and approximated critical values
both perform well and control the type I error rates.
While for the asymptotic test with the normal approximation in (\ref{t2_convergency}),
it does not provide a good performance when $n$ is small.
In particular when $n=5$, the type I error rate will be nearly zero so that the test will be extremely conservative.
Under scenario $\mathcal{S}_3$, given that the asymptotic test is not available,
we thus report the type I error rates for the tests with the exact and approximated critical values only.
Both tests control the type I error rates
and they perform equally well for any fixed sample size.

\subsection{Statistical power}\label{statistical_power}
In this section, we assess the ability of the three tests for detecting the skewness when the alternative distribution is skewed.
For this purpose, we reconsider the four normal-related distributions in Section \ref{section_motivation},
Skew-normal$(0,1,-10)$, Half-normal$(0,1)$, Log-normal$(0,1)$ and 0.3*$N(-2,1)$+0.7*$N(2,1)$,
as the alternative distributions for all the tests under different scenarios.
Then to numerically compute the statistical power,
we follow the same procedure as in Section \ref{t1e}
except that the sample data are now generated from the alternative distributions
rather than the standard normal distribution.
In addition, since the asymptotic test is suboptimal and the other two tests perform nearly the same,
we report the simulated power only for the tests with the approximated critical values in Figure \ref{fig:power} based on 1,000,000 simulations.

For the test under scenario $\mathcal{S}_1$, we note that it is always very powerful,
in particular for the three unimodal alternative distributions.
This is mainly because the extreme order statistics, including the minimum and maximum values,
are very sensitive to the tail behavior of the underlying distribution.
In contrast, the intermediate order statistics, including the first and third quartiles,
behave more stably and are less affected by the tail distributions \cite{Gather1995}.
As a consequence, the test under scenario $\mathcal{S}_2$ is often less powerful in detecting the skewness of data, as those reflected in the power curves for the three unimodal alternative distributions.
But there are also exceptional cases. 
Specifically, for the mixture-normal distribution,
since the two tails are both normally shaped, the minimum and maximum values behave similarly
so that the mid-range, $(a+b)/2$, is quite stable along with the sample size,
and consequently, it diminishes the ability of detecting the skewness.
On the other side, we note that the median is closer to the third quartile rather than to the first quartile,
and so the test under scenario $\mathcal{S}_2$ turns out to be more powerful than
the test under scenario $\mathcal{S}_1$ in the mixture-normal case.
Finally for the test under scenario $\mathcal{S}_3$, since it takes into account both the extreme and intermediate order statistics,
it is not surprising that it always performs better than, or at least as well as, the other two tests in most settings.

To sum up, by virtue of the well-controlled type I error rates and the reasonable statistical power,
we believe that our easy-to-implement tests with the approximated critical values will have potential to be widely adopted for detecting the skewness away from normality based on the five-number summary with application to meta-analysis.

\subsection{Simulated meta-analysis}\label{simulated_meta}
\noindent
To further demonstrate the usefulness of the proposed skewness tests, 
we also conduct a simulated meta-analysis consisting of 10 studies with normal data 
and 5 studies with non-normal data. 
Following the random-effects model \cite{higgins2008}, we first generate the individual means $\mu_i$, $i=1,2,\ldots,15$, from the between-study distribution $N(\mu, \tau^2)$.
Then for each study, we generate a sample of size $n_i$ from $N(\mu_i, \sigma_i^2)$ for $i=1,\dots,10$, and from Skew-normal$(\delta_i,\omega_i,\alpha_i)$ for $i=11,\dots, 15$, where $\delta_i=\mu_i-\omega_i\sqrt{2\alpha_i^2/[\pi(1+\alpha_i^2)]}$ ensuring that the mean of the skew-normal distribution is also $\mu_i$. 
Moreover, for the first 10 studies, we follow a similar setting as in Brockwell and Gordon\cite{Brockwell2001} 
and consider $\mu=0.5$, $\tau^2=0.04$, $\sigma_i^2=1$ for $i=1,\dots,10$, and 
$n_i=n$ for all the studies. 
While for the last 5 studies, we let  $\omega_i=5$ for $i=11,\dots,15$, and $\alpha_{11}=-0.1$ and $\alpha_i=-10$ for $i=12,\dots,15$ to represent the different levels of skewness. 
Lastly, it is noteworthy that we have also considered more general settings including unequal within-study variances and unequal sample sizes, and the comparison results remain similar. 

\begin{figure}[H]
	\centering
	\includegraphics[width=0.7\textwidth]{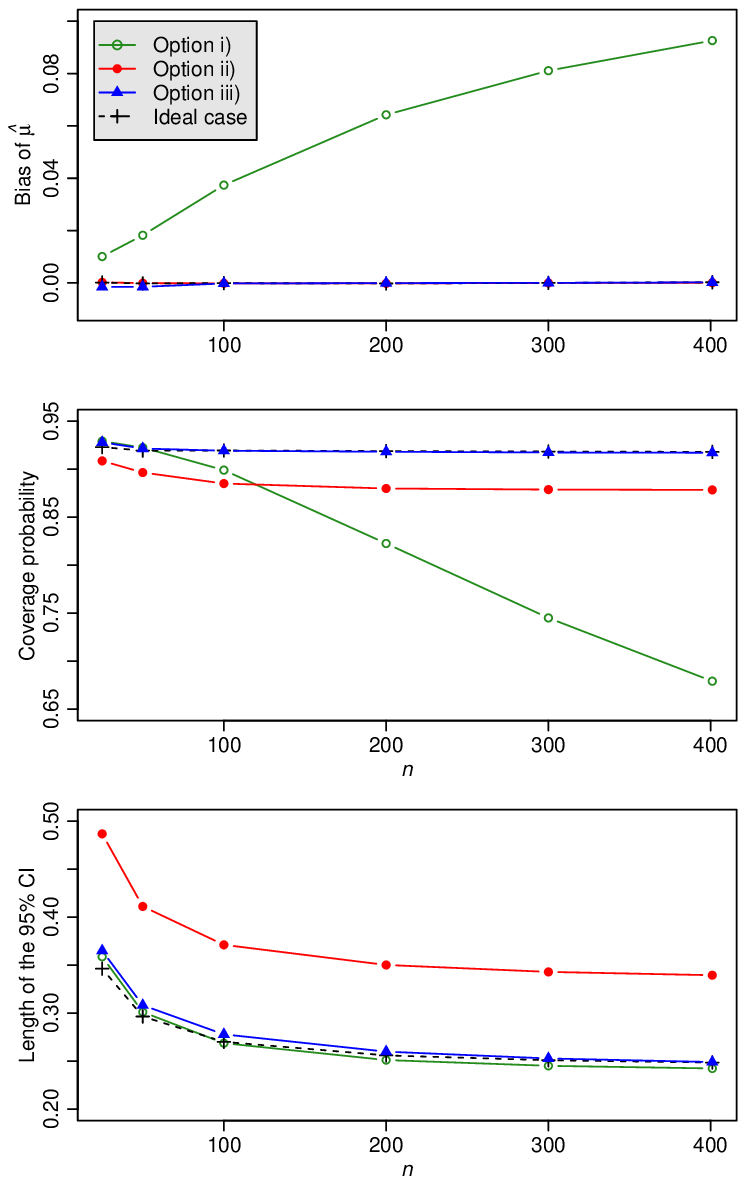}
	\setlength{\abovecaptionskip}{6pt}%
	\setlength{\belowcaptionskip}{0pt}%
	\caption{The bias of $\hat{\mu}$, the coverage probability and the average length for the  95\% CI for $\mu$ under scenario $\mathcal{S}_1$ with $n$ up to 401, where option i) represents that all the 15 studies are included, option ii) represents that only the first 5 studies reporting the sample mean and SD are included,  
		option iii) represents that the first 5 studies plus all other studies passing the skewness test are included,
		and the ideal case represents that the 10 normal studies are included with their true sample means and SDs.}\label{fig:simulated1}
\end{figure}

After the dataset is generated, 
we report the sample mean and SD as the summary statistics for the first 5 studies,
but instead report the minimum, median, and maximum values, i.e., under scenario $\mathcal{S}_1$,
for the other 10 studies.
We further consider three options to carry out the meta-analysis using i) all the 15 studies,
ii) only the first 5 studies reporting the sample mean and SD, and
iii) the first 5 studies plus all other studies passing the skewness test.
Moreover, as an ideal case for comparison, we also conduct the meta-analysis based on the true sample means and SDs of the generated data from the 10 normal studies. 
Finally, by considering the mean value as the effect size, we apply the DerSimonian and Laird method \cite{DerSimonian_Laird} for meta-analysis and report the bias of the effect size estimate $\hat\mu$, the coverage probability and the average length of the 95\% confidence interval (CI) for $\mu$, in Figure \ref{fig:simulated1} for the sample size up to 401 based on a total of 500,000 simulations.

From the top panel of Figure \ref{fig:simulated1},
it is evident that the effect size estimate $\hat\mu$ under option i) with all 15 studies being included tends to be significantly biased. 
In contrast, options ii) and iii) are both able to control the estimation bias of $\hat\mu$, nearly as well as the ideal case for benchmarking. 
From the middle and bottom panels of Figure 6, we also observe that option ii) not only suffers from a lower coverage probability,
but also has a wider CI compared to option iii) and the ideal case.
This is mainly because option ii) loses valuable information by excluding the normal studies reported with the five-number summary, and consequently yields less efficient estimation of the effect size. 
Taken together, our new option iii) with the flow chart in Figure \ref{fig:meta_procedure} can  effectively detect and exclude some very skewed studies away from the subsequent meta-analysis, and as seen from the simulation results, 
it performs equally well as the ideal case except that the average length of the CI is slightly longer. 

Finally, to save space, we present the simulation results under scenarios $\mathcal{S}_2$ and $\mathcal{S}_3$ 
in Appendix B; for more details, see Figures \ref{fig:simulated2} and \ref{fig:simulated3}. 
It is evident that the comparison results under scenario $\mathcal{S}_3$ are similar to those under scenario $\mathcal{S}_1$.
While for scenario $\mathcal{S}_2$, by noting that the test statistic $T_2$ is less powerful in detecting the skewness of data as clarified in Section \ref{statistical_power} through Figure \ref{fig:power}, 
option iii) may not be able to exclude some very skewed studies from the meta-analysis. 
As a consequence, it may also yield a biased effect size estimate, 
even though it is apparently better than option i). 
On the other hand, thanks to the flow chart in Figure \ref{fig:meta_procedure} 
that allows more studies to be included in the meta-analysis, 
option iii) is able to provide a narrower or much narrower CI compared to option ii). 
To summarize, although option iii) does not perform equally well as the ideal case under scenario $\mathcal{S}_2$, 
it is still comparable to, or better than, the other two options based on our simulation results.

\section{Real data analysis}\label{section_realdata}
To illustrate the usefulness of the skewness tests as well as the flow chart,
we now revisit the motivating example in Section \ref{section_motivation},
where Wu and Yang\cite{wuyang2020} investigated the impact of COVID-19 on the liver dysfunction.
Specifically, we first present their meta-analytical results
and then apply our new flow chart to reanalyze the example,
followed by a comparison made between their results and our new results.

\subsection{Original results in Wu and Yang\cite{wuyang2020}}
\begin{figure}[H]
	\centering
	\includegraphics[width=0.9\textwidth]{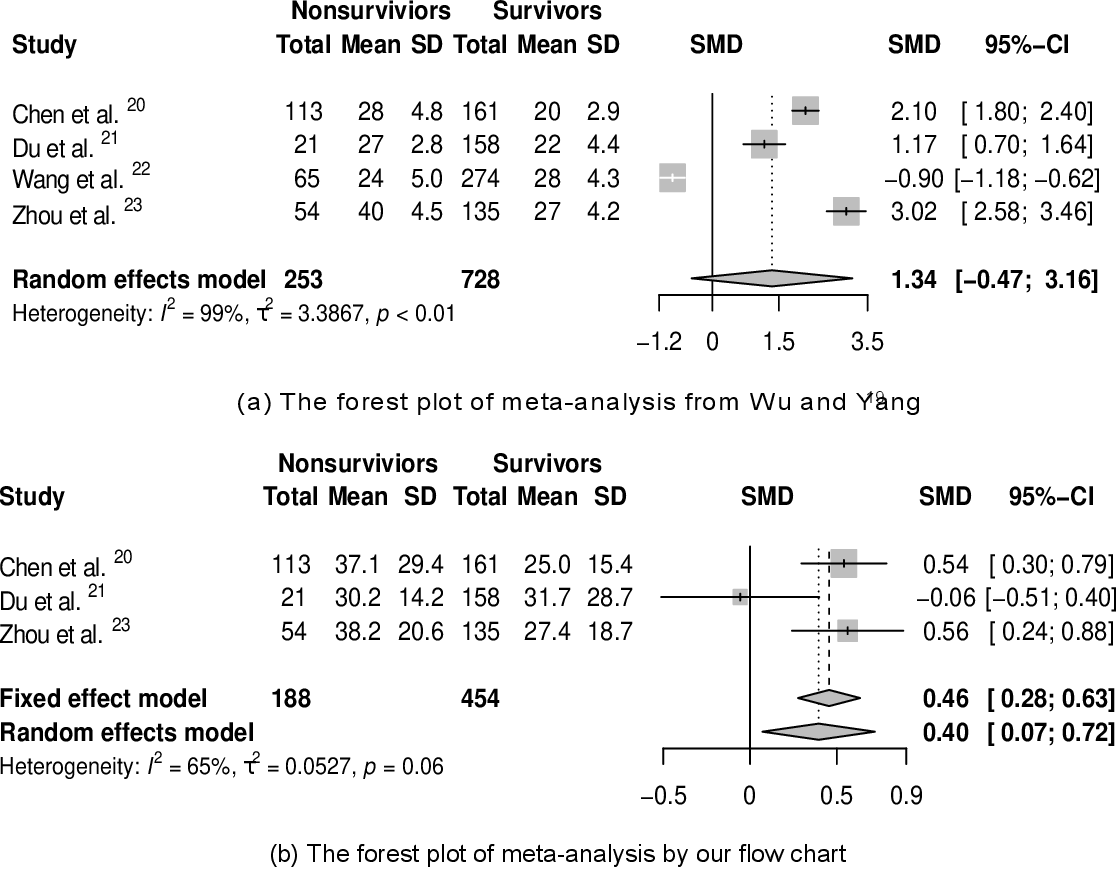}
	\setlength{\abovecaptionskip}{6pt}%
	\setlength{\belowcaptionskip}{0pt}%
	\caption{The forest plots of the meta-analyses from Wu and Yang\cite{wuyang2020} and by our new flow chart are presented in panels (a) and (b), respectively.}\label{figcase_new_1}
\end{figure}

To deal with the studies reported with the first quartile, the median and the third quartile
together with the sample size,
Wu and Yang\cite{wuyang2020} applied the data transformation in Hozo et al.\cite{hozo2005} to obtain the sample mean and SD estimates,
and then performed the meta-analysis with the forest plot in panel (a) of Figure \ref{figcase_new_1}.
The random-effects model was used to pool the studies,
which yielded the overall standardized mean difference (SMD) 1.34 with the 95\% CI being $[-0.47,3.16]$.
Given that the 95\% CI of the overall effect size covers zero,
Wu and Yang\cite{wuyang2020} concluded that
the impact of COVID-19 on the ALT level is not statistically significant.
However, noting that an extremely large heterogeneity with $p<0.01$ and $I^2=99\%$ is observed,
the random-effects model may not be sufficient to well synthesize the included studies,
and thus the final conclusion can be problematic.

\subsection{New results by our flow chart}
To reanalyze the impact of COVID-19 on the liver dysfunction,
we follow the flow chart in Figure \ref{fig:meta_procedure} as a practical guideline for meta-analysis,
where the first and foremost step is to identify whether some studies are significantly skewed.
Since the summary statistics for the four studies are all reported under scenario $\mathcal{S}_2$,
we thus apply the test statistic $T_2$ to conduct the skewness test.
Specifically, by Section \ref{scenario2},
we compute the absolute value of the observed $T_2$ for each data,
compare it with the approximated critical value at the significance level of 0.05,
and then report the test result in panel (a) of Table \ref{case2_result}.
It is surprising to see that only the last study passes the skewness test for both nonsurvivors and survivors groups, and consequently,
Luo et al.\cite{luo2016} and Wan et al.\cite{Tong2014} can be applied to this study for obtaining the sample mean and SD estimates.

Moreover, recall that the skewed data all display a positively skewed pattern as observed, and in the medical literature,
the log transformation is frequently used to normalize such data \cite{higgins2008skewed,Yamaguchi2017,shi2020b}.
Following this, we also apply a log transformation to the data from the three skewed studies, and then redo the skewness test using the log scale data.
From the test results in panel (b) of Table \ref{case2_result},
it is unfortunate that the nonsurvivors group in Wang et al.\cite{wang2020}, once again, fails to pass the skewness test, indicating an unacceptably large degree of skewness on the original data.

\begin{table}[h]\small
	\centering    
	\setlength{\abovecaptionskip}{4pt}%
	\setlength{\belowcaptionskip}{10pt}%
	\caption{The absolute values of the observed $T_2$, the critical values at the significance level of 0.05, and the test results for the original and log scale data from the corresponding studies are reported in panels (a) and (b), respectively.}\label{case2_result}
	\begin{tabular}{c|ccc|ccc}
		\multicolumn{7}{c}{(a) Test results for the original data}\\
		\hline
		\multirow{2}{*}{Study}            & \multicolumn{3}{c|}{Nonsurvivors}         & \multicolumn{3}{c}{Survivors}\\
		&$|T_2|$ &Critical value &Decision &$|T_2|$ &Critical value &Decision\\
		\hline
		Chen et al.\cite{chen2020}         &0.310   &0.249 &Reject	          &0.395  &0.209 &Reject\\
		Du et al.\cite{du2020}	      &0.176   &0.565 &Not reject	          &0.396  &0.211 &Reject\\		
		Wang et al.\cite{wang2020}         &0.667   &0.327 &Reject       &0.154  &0.160 &Not reject\\		
		Zhou et al.\cite{zhou2020}         &0.185   &0.359 &Not reject       &0.04  &0.228 &Not reject\\	
		\hline
		
	\end{tabular}	
	
	\vskip 10pt
	\begin{tabular}{c|ccc|ccc}
		\multicolumn{7}{c}{(b) Test results for the log scale data}\\
		\hline
		\multirow{2}{*}{Study}            & \multicolumn{3}{c|}{Nonsurvivors}         & \multicolumn{3}{c}{Survivors}\\
		&$|T_2|$ &Critical value &Decision &$|T_2|$ &Critical value &Decision\\
		\hline
		Chen et al.\cite{chen2020}         &0.083   &0.249 &Not reject	          &0.205  &0.209 &Not reject\\
		Du et al.\cite{du2020}	      &0.016   &0.565 &Not reject	          &0.151  &0.211 &Not reject\\		
		Wang et al.\cite{wang2020}         &0.495   &0.327 &Reject       &0.075  &0.160 &Not reject\\		
		\hline
		
	\end{tabular}
\end{table}

For the skewed studies, as per the proposed flow chart in Figure \ref{fig:meta_procedure},
one may exclude the skewed studies from meta-analysis
for normal data, or apply the non-normal data transformation methods for skewed studies,
or perform the subgroup analysis that separates the normal and skewed studies.
For this case, note that there are only four studies included
and further separating them into subgroups will make certain subgroup(s) 
include only one or two studies, which may not yield reliable results.
Therefore, we propose to take advantages of the first two options.
Specifically, we exclude Wang et al.\cite{wang2020} from the subsequent meta-analysis
given that its nonsurvivors group is extremely skewed and to the best of our knowledge, 
there is little work on directly meta-analyzing the skewed data with unknown distributions.
For both studies that pass the skewness test under the log scale (Chen et al.\cite{chen2020} and Du et al.\cite{du2020}),
we treat their original data as log-normally distributed
and estimate the sample means and SDs by Shi et al.\cite{shi2020b}

Then by following the setting in Wu and Yang\cite{wuyang2020},
we conduct the meta-analysis with the SMD as the effect size
and the forest plot is presented in panel (b) of Figure \ref{figcase_new_1}.
Noting that a moderate heterogeneity is observed with $I^2=65\%$ and $p=0.06$,
we refer to the random-effects model for the decision making \cite{higgins2008}.
Specifically, the random-effects model yields the SMD 0.40 with the 95\% CI being $[0.07,0.72]$,
which indicates that the nonsurvivors group has a significantly higher ALT level than the survivors group.
Meanwhile, it is also worth mentioning that if we exclude all skewed studies
but only include Zhou et al.\cite{zhou2020} in the meta-analysis according to our first option,
with the SMD being 0.56 and its 95\% CI being $[0.24,0.88]$,
the conclusion remains the same as above.
Consequently, we conclude that the nonsurvivors group has a significantly higher ALT level than the survivors group.

Of interest, the conclusion is converted
when we compare our new results with those in Wu and Yang.\cite{wuyang2020}
With the conflicting conclusions, it calls for more studies
to confirm the final conclusion so as to give a proper guideline in practice.
Meanwhile, more attention and methodologies are warranted
to deal with the skewed studies in meta-analysis.

\section{Conclusion}
For clinical studies with continuous outcomes,
the sample mean and standard deviation (SD) are routinely reported as the summary statistics
when the data are normally distributed.
While in some studies, however, researchers may report the whole or part of the five-number summary,
mainly because the data from the specific studies are potentially skewed away from normality.
For the studies with skewed data, if we include them in the classical meta-analysis for normal data,
it may yield unreliable or even misleading conclusions.
In this paper, we develop three new tests for detecting the skewness of data 
for the flow chart of the meta-analysis based on the sample size and the five-number summary.
If the skewness test is not rejected, we then apply the normal-based transformation methods to recover the sample mean and SD from the five-number summary.
Otherwise, we provide practitioners with three options for different cases.
Simulation studies are carried out to demonstrate that the
skewness tests yield satisfying statistical power with the type I error controlled.
The usefulness of the flow chart including the skewness tests has also been demonstrated
by the simulated meta-analysis as well as a real data example.
An online calculator is provided for performing the skewness test and the data transformation in the flow chart.
We also summarize the three skewness tests together with 
the critical regions of size 0.05 in Table \ref{test_statistics}.

\begin{table}[h]\small
	\setlength{\abovecaptionskip}{2pt}%
	\setlength{\belowcaptionskip}{6pt}%
	\caption{The summary table of the skewness tests under the three scenarios.} \label{test_statistics}
	\centering
	\begin{tabular}{ccc}
		\hline
		~~Scenario~~~            & ~~~~\text{Test statistic}                                 & ~~~~~~~~~\text{Critical region of size 0.05}~~~
		\\   \hline \\[-8pt]
		$\mathcal{S}_1$     & $T_1=\dfrac{a+b-2m}{b-a}$                  &$|T_1|>\dfrac{1}{\ln(n+9)}+\dfrac{2.5}{n+1}$
		\\  \hline \\[-8pt]
		$\mathcal{S}_2$     & $T_2=\dfrac{q_{1}+q_{3}-2m}{q_{3}-q_{1}}$  &$|T_2|>\dfrac{2.65}{\sqrt{n}}-\dfrac{6}{n^2}$
		\\  \hline \\[-8pt]
		$\mathcal{S}_3$     & $T_3=\max\left\{\dfrac{2.65\ln (0.6n)}{\sqrt{n}}|T_1|, |T_2|\right\}$                 &$T_3>\dfrac{3}{\sqrt{n}}-\dfrac{40}{n^3}$ \\ \\[-8pt]
		\hline
	\end{tabular}
\end{table}

To further clarify, if a study passes the skewness test,
it does not necessarily mean that the data are normally distributed, 
but rather there is no significant evidence to claim that the data are significantly skewed. And without further evidence (for or against) whether the studies passing the skewness test are truly normally distributed, our flow chart in Figure \ref{fig:meta_procedure} suggests to still include them into the subsequent meta-analysis. 
As otherwise, we will face another dilemma that valuable information may be excluded from meta-analysis so that the final conclusion is less reliable or even misleading, especially when a large proportion of studies are reported with the five-number summary. Future research, either theoretically or numerically, are warranted to further assess the studies passing the skewness test. 
Lastly, thanks to the good performance of the skewness tests as well as the flow chart, together with the online calculator at \url{http://www.math.hkbu.edu.hk/~tongt/papers/median2mean.html}, 
we expect they may have potential to be widely adopted in meta-analysis and evidence-based practice.

%\newpage
\

\section*{Appendix A: Theoretical results}
\begin{lemma}\label{lemma1}
	Let $Z_1,Z_2,\ldots,Z_n$ be a random sample from the standard normal distribution,
	and $Z_{(1)}\leq Z_{(2)}\leq \cdots \leq Z_{(n)}$ be the corresponding order statistics.
	We have
	\begin{equation*}
		(Z_{(1)},Z_{(2)},\ldots,Z_{(n)})\overset{d}{=}(-Z_{(n)},-Z_{(n-1)},\ldots,-Z_{(1)}),
	\end{equation*}
	where $\overset{d}{=}$ represents that two random vectors follow the same distribution.
	Thus it is evident that
	\begin{equation*}
		E(Z_{(i)})=-E(Z_{(n-i+1)}), \quad 1\leq i\leq n.
	\end{equation*}
	Specifically, by letting $n=4Q+1$ with $Q$ being a positive integer,
	it directly follows that
	$E(Z_{(2Q+1)})=0$, $E(Z_{(Q+1)})=-E(Z_{(3Q+1)})$ and $E(Z_{(1)})=-E(Z_{(n)})$.
	For a random sample $X_1,X_2,\ldots,X_n$ from the normal distribution with mean $\mu$ and variance $\sigma^2$,
	where $X_{(1)}\leq X_{(2)}\leq \cdots \leq X_{(n)}$ represents its order statistics, we have
	\begin{equation*}
		E(X_{(i)})=\mu+\sigma E(Z_{(i)}), \quad 1\leq i\leq n.
	\end{equation*}
\end{lemma}

\begin{theorem}\label{asy_s1}
	Under scenario $\mathcal{S}_1=\{a,m,b;n\}$, as $n\rightarrow \infty$ under the null hypothesis of normality,
	we have
	\begin{equation*}
		\sqrt{2\ln (n)}\xi(n)\left(\dfrac{a+b-2m}{b-a}\right) \overset{D}{\longrightarrow} {\rm Logistic}(0,1),
	\end{equation*}
	where $\xi(n)=2\Phi^{-1}[(n-0.375)/(n+0.25)]$.
\end{theorem}

\vskip 12pt
\noindent {\bf Proof}.
Under the null hypothesis of normality, it is evident that
\begin{equation*}
	\dfrac{a+b-2m}{2\sigma}=\dfrac{Z_{(1)}+Z_{(n)}}{2}-Z_{(2Q+1)}.
\end{equation*}
According to Ferguson\cite{Ferguson1996}, as $n\rightarrow \infty$, we have
\begin{equation*}
	\sqrt{2\ln (n)}\left(\dfrac{Z_{(1)}+Z_{(n)}}{2}\right)\overset{D}{\longrightarrow} {\rm Logistic}(0,0.5)
\end{equation*}
and
\begin{equation*}
	\sqrt{n}Z_{(2Q+1)}\overset{D}{\longrightarrow} N\left(0,\dfrac{\pi}{2}\right),
\end{equation*}
where $\overset{D}{\longrightarrow}$ denotes the convergence in distribution.
Thus under the null hypothesis, it is evident that
\begin{equation*}
	\sqrt{2\ln (n)}\left(\dfrac{a+b-2m}{2\sigma}\right)= I_1-I_2,
\end{equation*}
where
\begin{align*}
	I_1&=\sqrt{2\ln (n)}\left(\dfrac{Z_{(1)}+Z_{(n)}}{2}\right),\\
	I_2&=\sqrt{2\ln (n)}Z_{(2Q+1)}.
\end{align*}
As $n\rightarrow \infty$, $I_1\overset{D}{\longrightarrow} {\rm Logistic}(0,0.5)$
and $I_2\overset{P}{\longrightarrow}0$, where $\overset{P}{\longrightarrow}$ denotes the convergence in probability.
Then by Slutsky's Theorem, as $n\rightarrow \infty$, we conclude that
\begin{equation*}
	\sqrt{2\ln (n)}\left(\dfrac{a+b-2m}{\sigma}\right)\overset{D}{\longrightarrow} {\rm Logistic}(0,1).
\end{equation*}

According to Wan et al.\cite{Tong2014}, it is evident that
$\hat{\sigma}_1=(b-a)/\xi(n)$ is a consistent estimator of the standard deviation $\sigma$
for a normal distribution.
%Further by Lemma 1 in \cite{shi2018}, as $n\rightarrow \infty$, $\xi(n)=2\sqrt{2\ln (n)}$.
With $\sigma$ estimated by $\hat{\sigma}_1$ and noting that $\xi(n)$ is the fixed value for any given $n$, the final test statistic is derived as
\begin{equation*}\label{w2_s1}
	T_1=\dfrac{a+b-2m}{b-a}
\end{equation*}
and again by Slutsky's Theorem as $n\rightarrow \infty$,
$\sqrt{2\ln(n)}\xi(n)T_1\overset{D}{\longrightarrow} {\rm Logistic}(0,1)$.

\begin{theorem}\label{sampling_distribution_s1}
	Under scenario $\mathcal{S}_1=\{a,m,b;n\}$, the null distribution of the test statistic $T_1=(a+b-2m)/(b-a)$ is
	\begin{align}\label{sampling_distribution_s1_1}
		f_1(t_1)=&\iint\limits_{\mathcal{D}}\dfrac{n!}{[(2Q-1)!]^2}\cdot\dfrac{v-u}{2}\cdot\phi(u)\phi(v)\phi\left(\dfrac{u+v}{2}-\dfrac{t_1(v-u)}{2}\right)\nonumber \\
		&\cdot\left[\Phi\left(\dfrac{u+v}{2}-\dfrac{t_1(v-u)}{2}\right)-\Phi(u)\right]^{2Q-1}\nonumber\\
		& \cdot \left[\Phi(v)-\Phi\left(\dfrac{u+v}{2}-\dfrac{t_1(v-u)}{2}\right)\right]^{2Q-1}{\rm d}u{\rm d}v,
	\end{align}
	where
	%$\phi$ and $\Phi$ are the probability density function and the cumulative distribution function of the standard normal distribution, and
	$\mathcal{D}$ is the integral area that satisfies $u\leq v$.
	Furthermore, the null distribution of $T_1$ is symmetric about zero.
\end{theorem}

\vskip 12pt
\noindent {\bf Proof.}
Denote $z_1=Z_{(1)}$, $z_m=Z_{(2Q+1)}$ and $z_n=Z_{(n)}$.
Recall that for $(z_1,z_m,z_n)$, we have the joint distribution as
\begin{equation*}
	g_1(z_1,z_m,z_n)=\dfrac{n!}{[(2Q-1)!]^2}\phi(z_1)\phi(z_m)\phi(z_n)[\Phi(z_m)-\Phi(z_1)]^{2Q-1}[\Phi(z_n)-\Phi(z_m)]^{2Q-1}.
\end{equation*}
It is evident that $(z_1,z_m,z_n)\rightarrow(u,t_1,v)$ is a one-to-one mapping,
where $u=z_1$, $t_1=(z_1+z_n-2z_m)/(z_n-z_1)$ and $v=z_n$,
or equivalently, $z_1=u$, $z_m=(u+v)/2-t_1(v-u)/2$ and $z_n=v$.
Thus we have the determinant of the Jacobian matrix as
$$
\left |\begin{array}{cccc}
	1    &0   &0 \\
	\dfrac{1+t_1}{2} &\dfrac{u-v}{2} &\dfrac{1-t_1}{2}  \\
	0    &0   &1 \\
\end{array}\right|=\dfrac{v-u}{2}.
$$
With the Jacobian transformation, we derive the joint distribution of $(u,t_1,v)$ as
\begin{align*}\label{sampling_distribution_s1_1}
	g_1^*(u,t_1,v)=& \dfrac{n!}{[(2Q-1)!]^2}\cdot\dfrac{v-u}{2}\cdot\phi(u)\phi(v)\phi\left(\dfrac{u+v}{2}-\dfrac{t_1(v-u)}{2}\right)\nonumber \\
	&\cdot\left[\Phi\left(\dfrac{u+v}{2}-\dfrac{t_1(v-u)}{2}\right)-\Phi(u)\right]^{2Q-1}\nonumber \\
	&\cdot \left[\Phi(v)-\Phi\left(\dfrac{u+v}{2}-\dfrac{t_1(v-u)}{2}\right)\right]^{2Q-1}.
\end{align*}
Further by taking the integrals with respect to $u$ and $v$, we achieve the sampling distribution of $T_1$ in (\ref{sampling_distribution_s1_1}) under the null hypothesis.

In addition, by Lemma \ref{lemma1},
we have
\begin{equation*}
	(Z_{(1)},Z_{(2Q+1)},Z_{(n)})\overset{d}{=}(-Z_{(n)},-Z_{(2Q+1)},-Z_{(1)}).
\end{equation*}
Thus under the null hypothesis, it is evident that
\begin{equation*}
	T_1=\dfrac{Z_{(1)}+Z_{(n)}-2Z_{(2Q+1)}}{Z_{(n)}-Z_{(1)}} \quad {\rm and}\quad
	-T_1=\dfrac{(-Z_{(n)})+(-Z_{(1)})-2(-Z_{(2Q+1)})}{(-Z_{(1)})-(-Z_{(n)})},
\end{equation*}
and therefore the null distribution of $T_1$ is symmetric about zero.

\begin{theorem}\label{asy_s2}
	Under scenario $\mathcal{S}_2=\{q_1,m,q_3;n\}$, as $n\rightarrow \infty$ under the null hypothesis of normality,
	we have
	\begin{equation*}
		0.74\sqrt{n}\left(\dfrac{q_1+q_3-2m}{q_3-q_1}\right) \overset{D}{\longrightarrow} N(0,1).
	\end{equation*}
\end{theorem}

\vskip 12pt
\noindent {\bf Proof.}
Under the null hypothesis of normality, it is evident that
\begin{equation*}
	\dfrac{q_1+q_3-2m}{\sigma}= Z_{(Q+1)}+Z_{(3Q+1)}-2Z_{(2Q+1)}.
\end{equation*}
According to Ferguson\cite{Ferguson1996}, as $n\rightarrow \infty$, we have that
$(Z_{(Q+1)},Z_{(2Q+1)},Z_{(3Q+1)})$ follows asymptotically a tri-variate normal distribution
with mean vector $(\Phi^{-1}(0.25),0,\Phi^{-1}(0.75))$ and covariance matrix $\Sigma$,
where
$$\Sigma=\dfrac{1}{n}\left(
\begin{matrix}
	1.86 & 0.99  & 0.62 \\
	0.99 & \pi/2 & 0.99 \\
	0.62 & 0.99  & 1.86
\end{matrix}
\right).
$$
Then by the Delta method, as $n\rightarrow \infty$, we have
\begin{equation*}
	Z_{(Q+1)}+Z_{(3Q+1)}-2Z_{(2Q+1)}=(1,-2,1)\left(
	\begin{array}{c}
		Z_{(Q+1)}\\
		Z_{(2Q+1)}\\
		Z_{(3Q+1)}
	\end{array}
	\right)\overset{D}{\longrightarrow} N(0,3.32/n).
\end{equation*}
Therefore, we achieve the asymptotic normality under the null hypothesis that
\begin{equation*}
	0.55\sqrt{n}\left(\dfrac{q_1+q_3-2m}{\sigma}\right)\overset{D}{\longrightarrow} N(0,1).
\end{equation*}

According to Wan et al.\cite{Tong2014},
it is evident that $\hat{\sigma}_2=(q_3-q_1)/\eta(n)$ is a consistent estimator of
the standard deviation $\sigma$ for a normal distribution,
where $\eta(n)=2\Phi^{-1}[(0.75n-0.125)/(n+0.25)]$.
Further by Theorem 1 in Shi et al.\cite{shi2018}, as $n\rightarrow \infty$,
$\eta(n)=2\Phi^{-1}(0.75)= 1.35$.
With $\sigma$ estimated by $\hat{\sigma}_2$ and noting $\eta(n)$ is a constant,
we propose the test statistic as 
\begin{equation*}
	T_2=\dfrac{q_1+q_3-2m}{q_3-q_1}
\end{equation*}
and by Slutsky's Theorem as $n\rightarrow \infty$,
$0.74\sqrt{n}T_2 \overset{D}{\longrightarrow} \text{N}(0,1)$.

\begin{theorem}\label{sampling_distribution_s2}
	Under scenario $\mathcal{S}_2=\{q_1,m,q_3;n\}$, the null distribution of the test statistic $T_2=(q_1+q_3-2m)/(q_3-q_1)$ is
	\begin{align}\label{sampling_distribution_s2_1}
		f_2(t_2)=& \iint\limits_{\mathcal{D}}\dfrac{n!}{[Q!(Q-1)!]^2}\phi(x)\phi\left(\dfrac{x+y}{2}-\dfrac{t_2(y-x)}{2}\right)\phi(y)
		[\Phi(x)]^{Q}[1-\Phi(y)]^{Q}\nonumber \\
		&\cdot\left[\Phi\left(\dfrac{x+y}{2}-\dfrac{t_2(y-x)}{2}\right)-\Phi(x)\right]^{Q-1}\nonumber\\
		&\cdot \left[\Phi(y)-\Phi\left(\dfrac{x+y}{2}-\dfrac{t_2(y-x)}{2}\right)\right]^{Q-1}{\rm d}x{\rm d}y,
	\end{align}
	where $\mathcal{D}$ is the integral area that satisfies $x\leq y$.
	Furthermore, the null distribution of $T_2$ is symmetric about zero.
\end{theorem}

\vskip 12pt
\noindent {\bf Proof.}
Denote $z_{q_1}=Z_{(Q+1)}$, $z_m=Z_{(2Q+1)}$ and $z_{q_3}=Z_{(3Q+1)}$.
Recall that for $(z_{q_1},z_m,z_{q_3})$, we have the joint distribution as
\begin{align*}
	g_2(z_{q_1},z_m,z_{q_3})=&\dfrac{n!}{Q!(Q-1)!(Q-1)!Q!}\phi(z_{q_1})\phi(z_m)\phi(z_{q_3})
	[\Phi(z_{q_1})]^{Q}[\Phi(z_m)-\Phi(z_{q_1})]^{Q-1}\\
	&[\Phi(z_{q_3})-\Phi(z_m)]^{Q-1}[1-\Phi(z_{q_3})]^{Q}.
\end{align*}
It is evident that $(z_{q_1},z_m,z_{q_3})\rightarrow(x,t_2,y)$ is a one-to-one mapping,
where $x=z_{q_1}$, $t_2=(z_{q_1}+z_{q_3}-2z_m)/(z_{q_3}-z_{q_1})$ and $y=z_{q_3}$,
or equivalently, $z_{q_1}=x$, $z_m=(x+y)/2-t_2(y-x)/2$ and $z_{q_3}=y$.
Thus we have the determinant of the Jacobian matrix as
$$
\left |\begin{array}{cccc}
	1    &0   &0 \\
	\dfrac{1+t_2}{2} &\dfrac{x-y}{2} &\dfrac{1-t_2}{2}  \\
	0    &0   &1 \\
\end{array}\right|=\dfrac{y-x}{2}.
$$
With the Jacobian transformation, we derive the joint distribution of $(x,t_2,y)$ as
\begin{align*}
	g_2^*(x,t_2,y)=& \dfrac{n!}{[Q!(Q-1)!]^2}\phi(x)\phi\left(\dfrac{x+y}{2}-\dfrac{t_2(y-x)}{2}\right)\phi(y)
	[\Phi(x)]^{Q}[1-\Phi(y)]^{Q}\\
	&\cdot\left[\Phi\left(\dfrac{x+y}{2}-\dfrac{t_2(y-x)}{2}\right)-\Phi(x)\right]^{Q-1}\nonumber\\
	&\cdot \left[\Phi(y)-\Phi\left(\dfrac{x+y}{2}-\dfrac{t_2(y-x)}{2}\right)\right]^{Q-1}.
\end{align*}
Further by taking the integrals with respect to $x$ and $y$, we achieve the sampling distribution of $T_2$ in (\ref{sampling_distribution_s2_1}) under the null hypothesis.
Similar to the proof in Theorem \ref{sampling_distribution_s1},
we can prove that the null distribution of $T_2$ is symmetric about zero.

\begin{theorem}\label{k_n}
	Under scenario $\mathcal{S}_3=\{a,q_1,m,q_3,b;n\}$,
	the test statistic is derived as
	\begin{equation*}
		T_3=\max\left\{\dfrac{2.65\ln (0.6n)}{\sqrt{n}}\left|\dfrac{a+b-2m}{b-a}\right|,\ \left|\dfrac{q_1+q_3-2m}{q_3-q_1}\right|\right\}.
	\end{equation*}
\end{theorem}
\vskip 12pt
\noindent {\bf Proof.}
Under scenario $\mathcal{S}_3$, by taking advantages of both extreme and intermediate order statistics,
we consider to detect the skewness of data with
\begin{equation*}
	W_3=\max\left\{\left|\dfrac{a+b-2m}{{\rm SE}(a+b-2m)}\right|,\ \left|\dfrac{q_1+q_3-2m}{{\rm SE}(q_1+q_3-2m)}\right|\right\}.
\end{equation*}
Recall that in Section \ref{scenario1} of the main text, 
we have derived ${\rm SE}(a+b-2m)=\sigma \delta_1(n)$,
where $\delta_1(n)={\rm SE}(Z_{(1)}+Z_{(n)}-2Z_{(2Q+1)})$
and $\sigma$ is estimated with $(b-a)/\xi(n)$.
Similarly, we have ${\rm SE}(q_1+q_3-2m)=\sigma \delta_2(n)$,
where $\delta_2(n)={\rm SE}(Z_{(Q+1)}+Z_{(3Q+1)}-2Z_{(2Q+1)})$,
and $\sigma$ is estimated with $(q_3-q_1)/\eta(n)$.
Then, the test statistic is specified as
$$\max\left\{\dfrac{\xi(n)}{\delta_1(n)}|T_1|,\ \dfrac{\eta(n)}{\delta_2(n)}|T_2|\right\},$$
where $T_1=(a+b-2m)/(b-a)$ and $T_2=(q_1+q_3-2m)/(q_3-q_1)$
are the test statistics under scenarios $\mathcal{S}_1$ and $\mathcal{S}_2$.

To simplify the presentation, we combine $\eta(n)/\delta_2(n)$ with $\xi(n)/\delta_1(n)$ in the first term 	so that only one coefficient $k(n)=[\xi(n)/\delta_1(n)]/[\eta(n)/\delta_2(n)]$ is needed,
and meanwhile $k(n)|T_1|$ and $|T_2|$ are still comparable in scale.
Further noting that $k(n)$ is difficult to compute since $\delta_1(n)$ and $\delta_2(n)$
involved do not have the explicit forms,
the test statistic may not be readily accessible to practitioners.
By following the asymptotic form of $k(n)$,
we have provided the approximate formula as $k(n)\approx 2.65\ln (0.6n)/\sqrt{n}$ for practical use.
Finally, test statistic is yielded as
\begin{equation*}
	T_3=\max\left\{\dfrac{2.65\ln (0.6n)}{\sqrt{n}}\left|\dfrac{a+b-2m}{b-a}\right|,\ \left|\dfrac{q_1+q_3-2m}{q_3-q_1}\right|\right\}.
\end{equation*}

\newpage

\section*{Appendix B: Supplementary tables and figures}
\begin{figure}[H]
	\centering
	\includegraphics[width=0.75\textwidth]{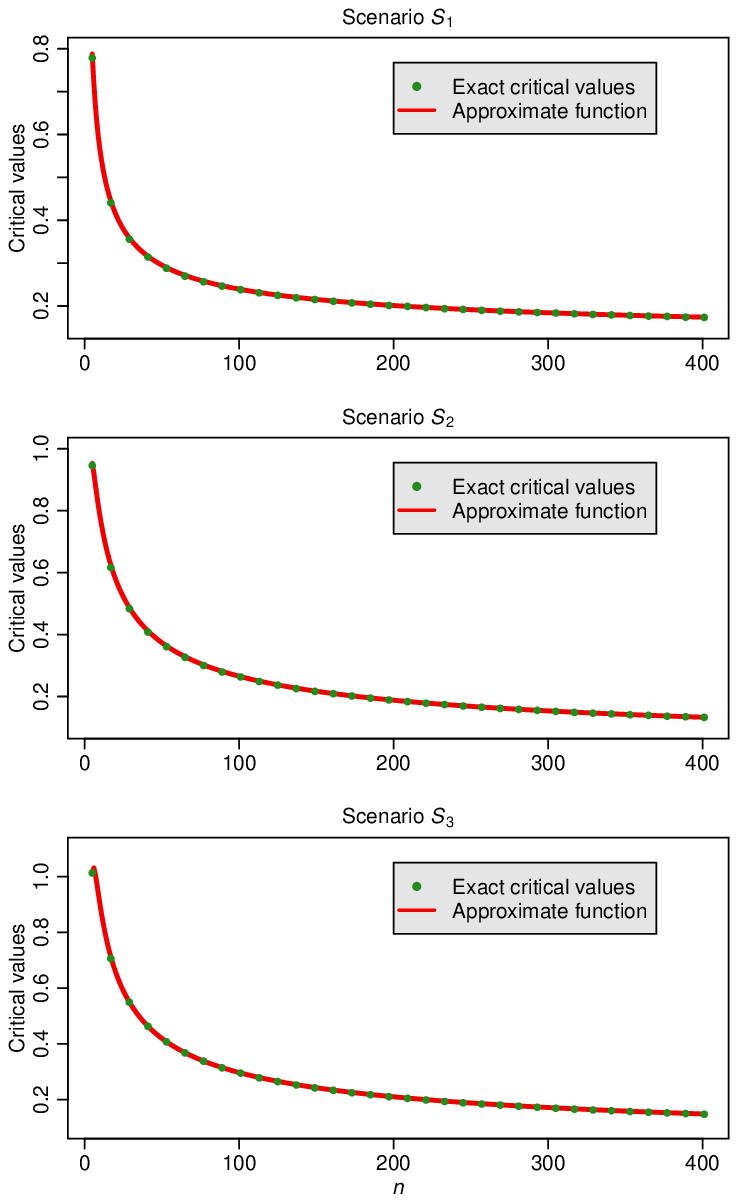}
	\caption{The green points represent the exact critical values under scenarios $\mathcal{S}_1$, $\mathcal{S}_2$ and $\mathcal{S}_3$, and the red lines represent the approximate functions of the critical values for $n$ up to 401.}
	\label{fig:c_combine}
\end{figure}

This appendix presents some supplementary tables and figures.
Specifically, Tables \ref{c1_table}-\ref{c3_table} report the numerical critical values of the test statistics in Section \ref{main}  under the three scenarios at the significance level of 0.05, Figure \ref{fig:c_combine} presents the approximate functions of the critical values under the three scenarios, 
and Figures \ref{fig:simulated2} and \ref{fig:simulated3} present the 
simulated meta-analysis results under scenarios $\mathcal{S}_2$ and $\mathcal{S}_3$, respectively.

For non-integer $Q$, the numerical critical values can be computed using the interpolation method through the following formula: 
$(1+[Q]-Q)c_{i,0.025}(4[Q]+1)+(Q-[Q])c_{i,0.025}(4[Q+1]+1)$ for $i=1,2,3$ respectively, 
where $[Q]$ represents the integer part of $Q$.
As an example, we now consider scenario $\mathcal{S}_1$ with $n=6$ and thus $Q=1.25$. 
By Table \ref{c1_table}, the critical values for the two adjacent integers 
$Q=1$ and $Q=2$ are 0.7792 and 0.5706, respectively. 
Further by the interpolation formula, the critical value for $Q=1.25$ can be computed as 0.72705 ($=0.75*0.7792+0.25*0.5706$). 
In addition, we can also apply the approximate formula $1/\ln (n+9)+2.5/(n+1)$  
in Section \ref{scenario1} to compute the critical value for $n=6$, 
yielding a value of 0.72641 which is very close to the interpolated value at 0.72705.

\begin{table}[H]\footnotesize
	\setlength{\abovecaptionskip}{0pt}%
	\setlength{\belowcaptionskip}{6pt}%
	\caption{The numerical values of $c_{1,0.025}(n)$ for $1\leq Q\leq100$, where $n=4Q+1$.}\label{c1_table}
	\centering
	\begin{tabular}{cc|cc|cc|cc|cc}
		\hline
		$Q$ &$c_{1,0.025}(n)$	&$Q$ &$c_{1,0.025}(n)$	   &$Q$ &$c_{1,0.025}(n)$	   &$Q$ &$c_{1,0.025}(n)$ &$Q$ &$c_{1,0.025}(n)$\\
		\hline
		1	&0.7792		    &21	&0.2505		&41	&0.2094     &61 &0.1920  &81 &0.1805 \\
		2	&0.5706		    &22	&0.2464		&42	&0.2087     &62 &0.1905  &82 &0.1803 \\
		3	&0.4964		    &23	&0.2433		&43 &0.2072     &63 &0.1903  &83 &0.1802 \\
		4	&0.4413		    &24	&0.2402		&44 &0.2067     &64 &0.1898  &84 &0.1794  \\
		5	&0.4032		    &25	&0.2375		&45 &0.2051     &65 &0.1892  &85 &0.1792 \\
		6	&0.3763		    &26	&0.2352		&46 &0.2042     &66 &0.1886  &86 &0.1786  \\
		7	&0.3554		    &27	&0.2332		&47 &0.2031     &67 &0.1878  &87 &0.1780\\
		8	&0.3395		    &28 &0.2315		&48 &0.2024     &68 &0.1877  &88 &0.1778 \\
		9	&0.3253		    &29	&0.2286		&49 &0.2013     &69 &0.1867  &89 &0.1777  \\
		10	&0.3132		    &30	&0.2277		&50 &0.2000     &70 &0.1864  &90 &0.1765    \\
		11  &0.3045          &31	&0.2243      &51 &0.1990     &71 &0.1858  &91 &0.1763 \\
		12	&0.2956          &32	&0.2238      &52 &0.1989     &72 &0.1850  &92 &0.1762 \\
		13	&0.2884          &33	&0.2219      &53 &0.1979     &73 &0.1848  &93 &0.1758 \\
		14	&0.2812          &34	&0.2203      &54 &0.1974     &74 &0.1840  &94 &0.1757 \\
		15	&0.2755          &35	&0.2183      &55 &0.1964     &75 &0.1837  &95 &0.1751 \\
		16	&0.2708          &36	&0.2172      &56 &0.1949     &76 &0.1836  &96 &0.1747 \\
		17	&0.2660          &37	&0.2151      &57 &0.1946     &77 &0.1823  &97 &0.1741 \\
		18	&0.2613          &38	&0.2135      &58 &0.1938     &78 &0.1819  &98 &0.1740 \\
		19	&0.2564          &39	&0.2128      &59 &0.1928     &79 &0.1818  &99 &0.1739 \\
		20	&0.2535          &40	&0.2111      &60 &0.1922     &80 &0.1811  &100 &0.1735 \\
		\hline
	\end{tabular}
\end{table}

\newpage
\begin{table}[H]\footnotesize
	\setlength{\abovecaptionskip}{0pt}%
	\setlength{\belowcaptionskip}{6pt}%
	\caption{The numerical values of $c_{2,0.025}(n)$ for $1\leq Q\leq100$, where $n=4Q+1$.}\label{c2_table}
	\centering
	\begin{tabular}{cc|cc|cc|cc|cc}
		\hline
		$Q$ &$c_{2,0.025}(n)$	&$Q$ &$c_{2,0.025}(n)$	   &$Q$ &$c_{2,0.025}(n)$	   &$Q$ &$c_{2,0.025}(n)$ &$Q$ &$c_{2,0.025}(n)$\\
		\hline
		1	&0.9463		    &21	&0.2861	&41	&0.2067	&61	&0.1692 &81	&0.1471 \\
		2	&0.8000		    &22	&0.2809	&42	&0.2034	&62	&0.1681 &82	&0.1461  \\
		3	&0.6913		    &23	&0.2748	&43 &0.2019	&63	&0.1667 &83	&0.1452  \\
		4	&0.6163		    &24	&0.2685	&44 &0.1993	&64	&0.1653 &84	&0.1443  \\
		5	&0.5594		    &25	&0.2633	&45 &0.1975	&65	&0.1641 &85	&0.1437  \\
		6	&0.5177		    &26	&0.2588	&46 &0.1954	&66	&0.1627 &86	&0.1428  \\
		7	&0.4819		    &27	&0.2538	&47 &0.1936	&67	&0.1614 &87	&0.1419  \\
		8	&0.4534		    &28 &0.2494	&48 &0.1914	&68	&0.1602 &88	&0.1409  \\
		9	&0.4297		    &29	&0.2447	&49 &0.1897	&69	&0.1593 &89	&0.1405  \\
		10	&0.4084		    &30	&0.2403	&50 &0.1879	&70	&0.1583 &90	&0.1395 \\
		11  &0.3903          &31	&0.2361  &51 &0.1854	&71	&0.1570   &91	&0.1391  \\
		12	&0.3744          &32	&0.2339  &52 &0.1831	&72	&0.1561   &92	&0.1381  \\
		13	&0.3608          &33	&0.2298  &53 &0.1823 	&73	&0.1551   &93	&0.1376  \\
		14	&0.3486          &34	&0.2267  &54 &0.1804 	&74	&0.1538   &94	&0.1364  \\
		15	&0.3372          &35	&0.2233  &55 &0.1785	&75	&0.1527   &95	&0.1357  \\
		16	&0.3266          &36	&0.2204  &56 &0.1776	&76	&0.1518   &96	&0.1355  \\
		17	&0.3179          &37	&0.2176  &57 &0.1757	&77	&0.1506   &97	&0.1345  \\
		18	&0.3085          &38	&0.2148  &58 &0.1749	&78	&0.1496   &98	&0.1339  \\
		19	&0.2999          &39	&0.2112  &59 &0.1721	&79	&0.1486   &99	&0.1332  \\
		20	&0.2931          &40	&0.2080  &60 &0.1718	&80	&0.1479  &100	&0.1326 \\
		\hline
	\end{tabular}
\end{table}

\begin{table}[H]\footnotesize
	\setlength{\abovecaptionskip}{0pt}%
	\setlength{\belowcaptionskip}{6pt}%
	
	\caption{The numerical values of $c_{3,0.05}(n)$ for $1\leq Q\leq100$, where $n=4Q+1$.}\label{c3_table}
	\centering
	\begin{tabular}{cc|cc|cc|cc|cc}
		\hline
		~$Q$ &~$c_{3,0.05}(n)$~	&~$Q$ &~$c_{3,0.05}(n)$~	   &~$Q$ &~$c_{3,0.05}(n)$~	   &~$Q$ &~$c_{3,0.05}(n)$~ &~$Q$ &~$c_{3,0.05}(n)$~\\
		\hline
		1	&1.0129		        &21	&0.3214 	&41	&0.2305 	&61	&0.1885  &81	&0.1635  \\
		2	&0.9062 		    &22	&0.3139 	&42	&0.2271 	&62	&0.1871  &82	&0.1626   \\
		3	&0.7929 		    &23	&0.3067 	&43 &0.2247 	&63	&0.1856  &83	&0.1617  \\
		4	&0.7060 		    &24	&0.3004 	&44 &0.2223 	&64	&0.1840  &84	&0.1607   \\
		5	&0.6416 		    &25	&0.2948 	&45 &0.2193 	&65	&0.1827  &85	&0.1600   \\
		6	&0.5898 		    &26	&0.2885 	&46 &0.2173 	&66	&0.1813  &86	&0.1587   \\
		7	&0.5490 		    &27	&0.2831 	&47 &0.2149 	&67	&0.1802  &87	&0.1579   \\
		8	&0.5151 		    &28 &0.2781 	&48 &0.2129 	&68	&0.1786  &88	&0.1570   \\
		9	&0.4870 		    &29	&0.2738 	&49 &0.2104 	&69	&0.1775  &89	&0.1561  \\
		10	&0.4630		        &30	&0.2687	    &50 &0.2082	    &70	&0.1762  &90	&0.1556 \\
		11  &0.4419           &31	&0.2645   &51 &0.2065 	&71	&0.1747    &91	&0.1546   \\
		12	&0.4229           &32	&0.2604   &52 &0.2043 	&72	&0.1734    &92	&0.1537   \\
		13	&0.4071           &33	&0.2564   &53 &0.2024  	&73	&0.1724    &93	&0.1528   \\
		14	&0.3929           &34	&0.2523   &54 &0.2004  	&74	&0.1713    &94	&0.1522   \\
		15	&0.3797           &35	&0.2489   &55 &0.1986 	&75	&0.1700    &95	&0.1512   \\
		16	&0.3675           &36	&0.2456   &56 &0.1971 	&76	&0.1689    &96	&0.1505   \\
		17	&0.3569           &37	&0.2419   &57 &0.1953 	&77	&0.1679    &97	&0.1497   \\
		18	&0.3473           &38	&0.2393   &58 &0.1933 	&78	&0.1669    &98	&0.1489   \\
		19	&0.3380           &39	&0.2359   &59 &0.1920 	&79	&0.1657    &99	&0.1479   \\
		20	&0.3290           &40	&0.2330   &60 &0.1902	&80	&0.1646    &100	&0.1472 \\
		\hline
	\end{tabular}
\end{table}

\begin{figure}[H]
	\centering
	\includegraphics[width=0.7\textwidth]{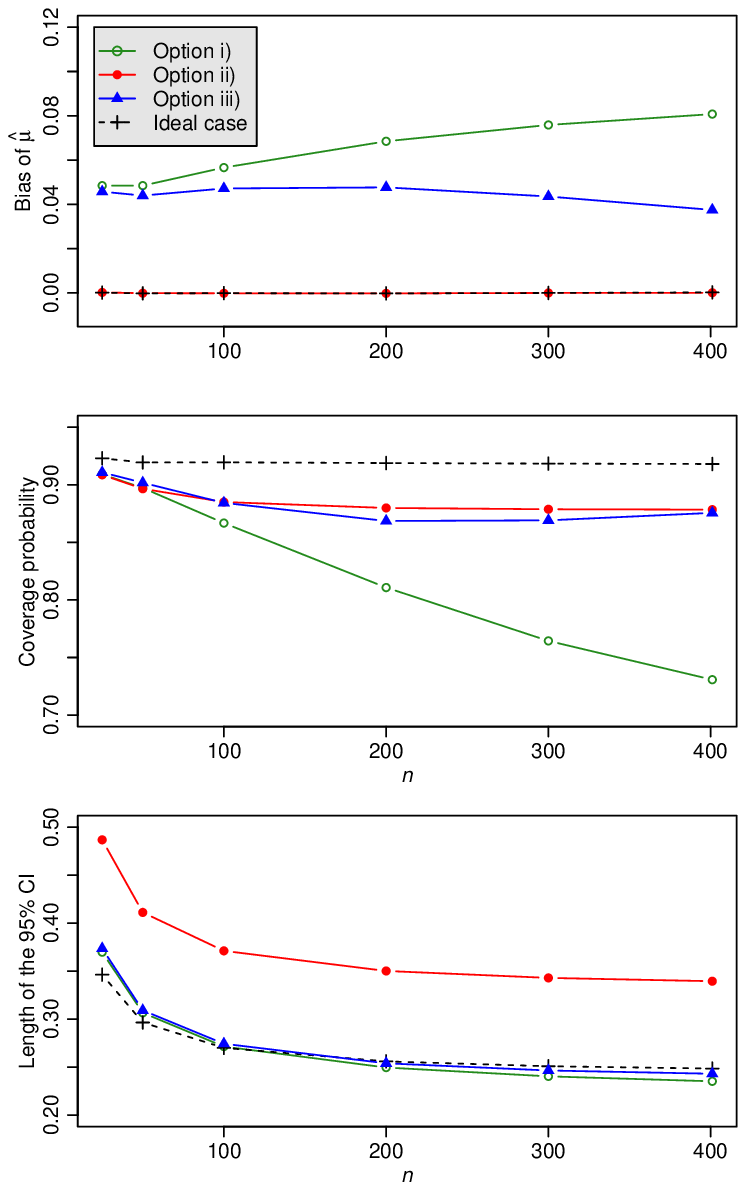}
	\setlength{\abovecaptionskip}{6pt}%
	\setlength{\belowcaptionskip}{0pt}%
	\caption{The bias of $\hat{\mu}$, the coverage probability and the average length of the  95\% CI for $\mu$ under scenario $\mathcal{S}_2$ with $n$ up to 401, where option i) represents that all the 15 studies are included, option ii) represents that only the first 5 studies reporting the sample mean and SD are included,  
		option iii) represents that the first 5 studies plus all other studies passing the skewness test are included,
		and the ideal case represents that the 10 normal studies are included with their  true sample means and SDs.}\label{fig:simulated2}
\end{figure}

\begin{figure}[H]
	\centering
	\includegraphics[width=0.7\textwidth]{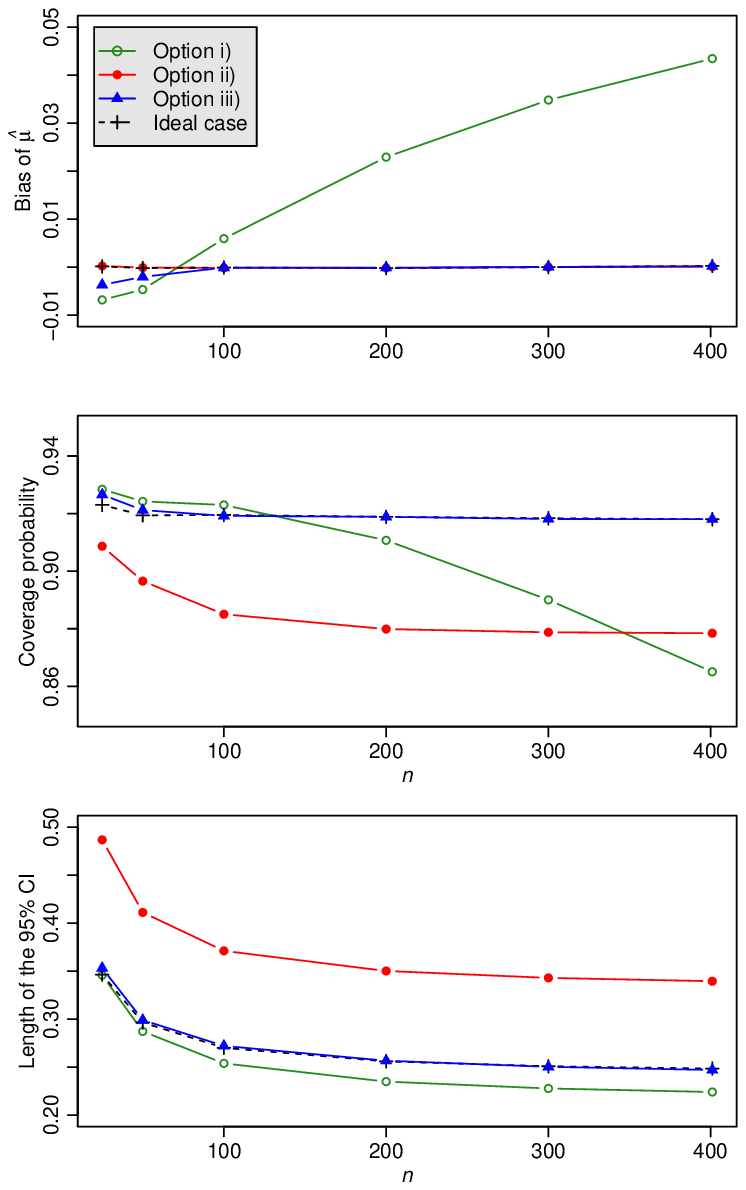}
	\setlength{\abovecaptionskip}{6pt}%
	\setlength{\belowcaptionskip}{0pt}%
	\caption{The bias of $\hat{\mu}$, the coverage probability and the average length of the  95\% CI for $\mu$ under scenario $\mathcal{S}_3$ with $n$ up to 401, where option i) represents that all the 15 studies are included, option ii) represents that only the first 5 studies reporting the sample mean and SD are included,  
		option iii) represents that the first 5 studies plus all other studies passing the skewness test are included,
		and the ideal case represents that the 10 normal studies are included with their  true sample means and SDs.}\label{fig:simulated3}
\end{figure}

\end{document}